\documentclass{article}%
\usepackage{amsfonts}
\usepackage{amsmath}
\usepackage{amssymb}
\usepackage{graphicx}
\usepackage{hyperref}%
\providecommand{\U}[1]{\protect\rule{.1in}{.1in}}
\begin{document}

\title{More on Rotations as Spin Matrix Polynomials}
\author{Thomas L. Curtright\smallskip\\{\small curtright@miami.edu\medskip}\\Department of Physics, University of Miami\\Coral Gables, FL 33124-8046, USA}
\date{}
\maketitle

\begin{abstract}
Any nonsingular function of spin $j$ matrices always reduces to a matrix
polynomial of order $2j$. \ The challenge is to find a convenient form for the
coefficients of the matrix polynomial. \ The theory of biorthogonal systems is
a useful framework to meet this challenge. \ Central factorial numbers play a
key role in the theoretical development. \ Explicit polynomial coefficients
for rotations expressed either as exponentials or as rational Cayley
transforms are considered here. \ Structural features of the results are
discussed and compared, and large $j$ limits of the coefficients are examined.

\end{abstract}

\section{Introduction}

As a consequence of the Cayley-Hamilton theorem \cite{Cayley,CayleyHamilton}
any nonsingular function \cite{Footnote1} of an $n\times n$ matrix always
reduces to a polynomial of order $n-1$ in powers of the matrix, since the
$n$th and higher powers of the matrix can be reduced to a linear combination
of lower powers (see \cite{Lehrer} and earlier literature cited therein).
\ The challenge for anyone wishing to take advantage of this fact is to find a
convenient form for the coefficients of the matrix polynomial. \ 

In particular, a nonsingular matrix function of any one component,
$\boldsymbol{\hat{n}\cdot J}$, for a unitary, irreducible angular momentum
representation of spin $j$, always reduces to a matrix polynomial of order
$2j$. \ Explicit polynomial coefficients for rotations expressed either as
exponentials or as rational Cayley transforms \cite{Cayley,CayleyTransform}
are considered here, to augment some recent studies \cite{CFZ,CvK,TSvK}.
\ Structural features of the results are discussed and compared, including the
behavior of the coefficients for large $j$.

The theory of biorthogonal systems \cite{Brody,CM,CMS} is a useful framework
to find the coefficients of the matrix polynomial, even in the most general
case. \ For analytic spin functions of $\boldsymbol{\hat{n}\cdot J}$, central
factorial numbers \cite{Riordan,CFN} play a key role in the construction of
the biorthogonal system. \ 

In prior work, Curtright, Fairlie, and Zachos (CFZ) obtained explicit and
intuitive results \cite{CFZ} expressing the standard, exponential rotation
matrix for \emph{any} quantized angular momentum $j$ as a polynomial of order
$2j$ in the corresponding $\left(  2j+1\right)  \times\left(  2j+1\right)  $
spin matrices $\boldsymbol{\hat{n}\cdot J}$ that generate rotations about
axis\ $\boldsymbol{\hat{n}}$. \ While many previous studies of this or closely
related problems can be found in the literature
\cite{Wigner,WignerAgain,Illamed,vanW,WW,T}, none of these other studies
succeeded to find such simple, compact expressions for the coefficients in the
spin matrix polynomial, as elementary functions of the rotation angle, as
those obtained by CFZ. \ For each angle-dependent coefficient in the
polynomial, the explicit formula found by CFZ involves nothing more
complicated than a truncated series expansion for a power of the $\arcsin$
function. \ \ Although a detailed proof of the CFZ result is not exhibited in
\cite{CFZ}, the essential ingredients needed to provide such a proof are in
that paper, and indeed, the details of two elementary derivations were
subsequently given in \cite{CvK}. \ 

More recently, Van Kortryk \cite{TSvK} discovered the corresponding polynomial
result for the Cayley rational form of an irreducible, unitary $SU(2)$
representation. \ At first glance the Cayley transform for a spin
representation would seem to follow immediately from the CFZ result just by
changing variables from $\theta$ to a function $\theta\left(  \alpha\right)
$, where $\alpha$ parameterizes the transform. \ For any given numerical value
of $\boldsymbol{\hat{n}\cdot J}$\ this would be so, obviously, but it is not
obviously so for matrix-valued $\boldsymbol{\hat{n}\cdot J}$. \ Nevertheless,
as it turns out, the result for the Cayley transform is actually much simpler
than the CFZ result for the exponential. \ For the Cayley transform the
explicit coefficients involve nothing more complicated than truncated series
expansions of already finite polynomials in a real parameter $\alpha$.
\ Remarkably, this is true for the Cayley transform of \emph{any} matrix, not
just spin matrices \cite{He,Householder,Faddeev}.

In this paper I compare the CFZ and Van Kortryk results, and I provide a more
complete discussion of the latter, including a detailed derivation of the
polynomial coefficients for the $SU(2)$ Cayley transforms. \ I show explicitly
how the CFZ results are connected to Cayley transforms by Laplace transformations.

\section{Methodology}

The theory that underlies construction of spin matrix polynomials is discussed
in several places in the literature (see \cite{CFZ,CvK} and references
therein). \ Here I condense and summarize the key results. \ 

\subsection{A fundamental identity}

The fundamental identity for spin $j\in\left\{  0,\frac{1}{2},1,\frac{3}%
{2},\cdots\right\}  $ is \cite{CvK}
\begin{equation}
\left(  \boldsymbol{\hat{n}\cdot J}\right)  ^{2j+1}=-\sum_{m=0}^{2j}t\left(
2+2j,1+m\right)  \times\left(  \boldsymbol{\hat{n}\cdot J}\right)  ^{m}\ ,
\label{FI}%
\end{equation}
where the coefficients $t\left(  n,k\right)  $\ are central factorial numbers
\cite{Riordan,CFN}. \ (For details, see Appendices A and B.) \ This identity
ensures that all powers of $\left(  \boldsymbol{\hat{n}\cdot J}\right)  $
higher than $2j$ can be reduced to linear combinations of lower powers. \ This
is a specific illustration of the Cayley-Hamilton theorem, applied to the
monomial on the LHS of (\ref{FI}).

\subsection{Vandermondeum}

The Vandermonde matrix \cite{Vandermonde} for spin $j$ is%
\begin{equation}
V\left[  j\right]  =\left[
\begin{array}
[c]{ccccc}%
1 & 2j & \left(  2j\right)  ^{2} & \cdots & \left(  2j\right)  ^{2j}\\
1 & 2j-2 & \left(  2j-2\right)  ^{2} & \cdots & \left(  2j-2\right)  ^{2j}\\
\vdots & \vdots & \vdots & \ddots & \vdots\\
1 & -2j & \left(  -2j\right)  ^{2} & \cdots & \left(  -2j\right)  ^{2j}%
\end{array}
\right]  \ , \label{Vandermonde}%
\end{equation}
while its inverse \cite{InverseVandermonde} is given by the \textquotedblleft
Fin du monde\textquotedblright\ matrix elements%
\begin{equation}
\left(  V^{-1}\left[  j\right]  \right)  _{kl}=\frac{2^{1-k}~\left(
-1\right)  ^{1-l}~N_{k}\left(  l,j\right)  }{\left(  2j+1-l\right)  !~\left(
l-1\right)  !}\ , \label{Findumonde}%
\end{equation}
where $k,l\in\left\{  1,2,\cdots,2j+1\right\}  $, and where the numerator
factors $N_{k}\left(  l,j\right)  $ are given by nested sums subject to some
exclusions.\
\begin{equation}
N_{k}\left(  l,j\right)  =\left(  -1\right)  ^{k-2j-1}\left\{
\begin{array}
[c]{ccc}%
1 & \text{for} & k=2j+1\\
&  & \\
\sum\limits_{\substack{1\leq m_{1}<m_{2}<\cdots<m_{2j+1-k}\leq2j+1\\m_{1}%
,m_{2},\cdots,m_{2j+1-k}\neq l}}\left(  j+1-m_{1}\right)  \cdots\left(
j+1-m_{2j+1-k}\right)  & \text{for} & 1\leq k<2j+1
\end{array}
\right.  \ . \label{Numerators}%
\end{equation}
These are unfamiliar polynomials \cite{Footnote2} in $l$ and $j$. \ For
$k=2j+1$ the numerator is always $1$, as it says above, and for $k=2j$ it
reduces to a single sum giving a result only first order in both $j$ and $l$.
\ Thus
\begin{equation}
N_{2j+1}\left(  l,j\right)  =1\ ,\ \ \ N_{2j}\left(  l,j\right)  =1-l+j\ .
\label{HighestIndexPolys}%
\end{equation}
But for other values of $k$ these numerator polynomials quickly get out of
hand. \ For example, for $k=2j-1$ the resulting double sum yields a polynomial
3rd order in $j$ and 2nd order in $l$.
\begin{equation}
N_{2j-1}\left(  l,j\right)  =\left(  1-l\right)  ^{2}+2\left(  1-l\right)
j+\frac{1}{6}\left(  1-j\right)  \left(  2j-1\right)  j\ .
\label{AnotherHighIndexPoly}%
\end{equation}

For reasonable values of $j$ --- up to $100$, say --- it is easiest to just
use numerical machine computation to obtain $V^{-1}\left[  j\right]  $
directly from $V\left[  j\right]  $, with no need of the analytic closed-form
expressions for the $N_{k}\left(  l,j\right)  $.

\subsection{Biorthogonal matrices}

It is useful to have in hand the \emph{dual matrices} which are trace
orthonormalized with respect to powers of the spin matrix, $S\equiv
2~\boldsymbol{\hat{n}\cdot J}$. \ Without loss of generality, choose
$S=2J_{3}$, since any other choice for $\boldsymbol{\hat{n}}$\ merely requires
selecting a different basis to diagonalize the spin matrix, thereby obtaining
the same eigenvalues as $2J_{3}$. \ Thus the powers are
\begin{equation}
S^{m}=\left[
\begin{array}
[c]{ccccc}%
\left(  2j\right)  ^{m} & 0 & \cdots & 0 & 0\\
0 & \left(  2j-2\right)  ^{m} & \cdots & 0 & 0\\
\vdots & \vdots & \ddots & \vdots & \vdots\\
0 & 0 & \cdots & \left(  -2j+2\right)  ^{m} & 0\\
0 & 0 & \cdots & 0 & \left(  -2j\right)  ^{m}%
\end{array}
\right]  \ .
\end{equation}
It is now straightforward to construct orthonormalized dual matrices $T_{n}$
such that%
\begin{equation}
\delta_{n,m}=\operatorname*{Trace}\left(  T_{n}~S^{m}\right)
\ ,\ \ \ n,m=0,1,\cdots,2j\ . \label{OrthoTrace}%
\end{equation}
The $T_{n}$\ may also be chosen to be diagonal $\left(  2j+1\right)
\times\left(  2j+1\right)  $ matrices in the basis that diagonalizes $S$. \ In
fact, for any spin $j$ the required entries on the diagonal of $T_{n}$ are
just the entries in the $\left(  n+1\right)  $st row of the inverted
Vandermonde matrix, $V^{-1}\left[  j\right]  $. \ (Note that here, unlike the
conventions in \cite{CFZ}, both rows and columns of the Vandermonde matrix and
its inverse are indexed as $1,2,\cdots,2j+1$.) \ That is, \
\begin{equation}
\left(  T_{n-1}\right)  _{kk}=\left(  V^{-1}\left[  j\right]  \right)
_{n,k}\ ,\ \ \ n,k=1,\cdots,2j+1\ . \label{DualMatrices}%
\end{equation}
This result follows immediately from the fact that the diagonal entries for
$S^{m}$ are just the entries in the corresponding column (i.e. the $\left(
m+1\right)  $st column) of the Vandermonde matrix, $V\left[  j\right]  $.

Explicit examples of spin matrix powers and their duals, and the corresponding
Vandermonde matrix inverse $V^{-1}\left[  j\right]  $, are given in Appendix C
for $j=1/2,\ 1,\ 3/2,$ and $2$.

\subsection{Projecting coefficients}

Suppose $f$ is a nonsingular matrix function for spin $j$. \ In a basis where
$S=2J_{3}$, the matrix $f\left(  S\right)  $ is diagonal and so is its
reduction to a polynomial in $S$:%
\begin{multline}
\left[
\begin{array}
[c]{ccccc}%
f\left(  2j\right)  & 0 & \cdots & 0 & 0\\
0 & f\left(  2j-2\right)  & \cdots & 0 & 0\\
\vdots & \vdots & \ddots & \vdots & \vdots\\
0 & 0 & \cdots & f\left(  -2j+2\right)  & 0\\
0 & 0 & \cdots & 0 & f\left(  -2j\right)
\end{array}
\right]  =f\left(  S\right) \nonumber\\
=\sum_{m=0}^{2j}f_{m}~S^{m}=\sum_{m=0}^{2j}f_{m}\left[
\begin{array}
[c]{ccccc}%
\left(  2j\right)  ^{m} & 0 & \cdots & 0 & 0\\
0 & \left(  2j-2\right)  ^{m} & \cdots & 0 & 0\\
\vdots & \vdots & \ddots & \vdots & \vdots\\
0 & 0 & \cdots & \left(  -2j+2\right)  ^{m} & 0\\
0 & 0 & \cdots & 0 & \left(  -2j\right)  ^{m}%
\end{array}
\right]  \ .
\end{multline}
Trace orthogonality of the dual matrices then gives the coefficients as%
\begin{equation}
f_{m}=\operatorname*{Trace}\left(  T_{m}~f\left(  S\right)  \right)  \ .
\end{equation}
Assembling the coefficients into a column and using (\ref{DualMatrices}) gives%
\begin{equation}
\left[
\begin{array}
[c]{c}%
f_{0}\\
f_{1}\\
\vdots\\
f_{2j-1}\\
f_{2j}%
\end{array}
\right]  =\left[
\begin{array}
[c]{c}%
\sum_{k=1}^{2j+1}\left(  V^{-1}\left[  j\right]  \right)  _{1,k}f\left(
2j+2-2k\right) \\
\sum_{k=1}^{2j+1}\left(  V^{-1}\left[  j\right]  \right)  _{2,k}f\left(
2j+2-2k\right) \\
\vdots\\
\sum_{k=1}^{2j+1}\left(  V^{-1}\left[  j\right]  \right)  _{2j,k}f\left(
2j+2-2k\right) \\
\sum_{k=1}^{2j+1}\left(  V^{-1}\left[  j\right]  \right)  _{2j+1,k}f\left(
2j+2-2k\right)
\end{array}
\right]  =V^{-1}\left[  j\right]  \left[
\begin{array}
[c]{c}%
f\left(  2j\right) \\
f\left(  2j-2\right) \\
\vdots\\
f\left(  2-2j\right) \\
f\left(  -2j\right)
\end{array}
\right]  \ . \label{coefficients via Findumonde}%
\end{equation}
So the spin matrix polynomial coefficients for $f\left(  S\right)  $ are
determined by straightforward matrix multiplication using the inverse
Vandermonde matrix. \ 

In terms of the explicit form in (\ref{Findumonde}), for $k=1,2,\cdots,2j+1$,
\begin{align}
f_{k-1}  &  =\sum_{l=1}^{2j+1}\left(  V^{-1}\left[  j\right]  \right)
_{kl}~f\left(  2j+2-2l\right)  =2^{1-k}~\sum_{l=1}^{2j+1}\frac{\left(
-1\right)  ^{1-l}~N_{k}\left(  l,j\right)  ~f\left(  2j+2-2l\right)  }{\left(
2j+1-l\right)  !~\left(  l-1\right)  !}\nonumber\\
&  =2^{1-k}~\sum_{l=1}^{2j+1}\frac{\left(  -1\right)  ^{k-l-2j}f\left(
2j+2-2l\right)  }{\left(  2j+1-l\right)  !~\left(  l-1\right)  !}%
\sum\limits_{\substack{1\leq m_{1}<\cdots<m_{2j+1-k}\leq2j+1\\m_{1}%
,m_{2},\cdots,m_{2j+1-k}\neq l}}\left(  j+1-m_{1}\right)  \cdots\left(
j+1-m_{2j+1-k}\right)  \ .
\end{align}
The coefficients of the highest powers appearing in the spin matrix polynomial
are often worked out readily from the general closed form expressions for the
$N_{k}\left(  l,j\right)  $, as given in (\ref{HighestIndexPolys}) and
(\ref{AnotherHighIndexPoly}). \ For the highest three powers in the
polynomial,
\begin{align}
f_{2j}  &  =2^{-2j}~\sum_{l=1}^{2j+1}\frac{\left(  -1\right)  ^{1-l}~f\left(
2j+2-2l\right)  }{\left(  2j+1-l\right)  !~\left(  l-1\right)  !}\ ,\\
f_{2j-1}  &  =2^{1-2j}~\sum_{l=1}^{2j+1}\frac{\left(  -1\right)
^{1-l}~\left(  1-l+j\right)  ~f\left(  2j+2-2l\right)  }{\left(
2j+1-l\right)  !~\left(  l-1\right)  !}\ ,\\
f_{2j-2}  &  =2^{2-2j}~\sum_{l=1}^{2j+1}\frac{\left(  -1\right)
^{1-l}~\left(  \left(  1-l\right)  ^{2}+2\left(  1-l\right)  j+\frac{1}%
{6}\left(  1-j\right)  \left(  2j-1\right)  j\right)  ~f\left(
2j+2-2l\right)  }{\left(  2j+1-l\right)  !~\left(  l-1\right)  !}\ ,
\end{align}
etc. \ But again, for reasonable values of $j$ it is probably easiest to just
use numerical machine computation to first obtain $V^{-1}\left[  j\right]  $
directly from $V\left[  j\right]  $, and then perform the matrix
multiplication in (\ref{coefficients via Findumonde}).

Indeed, by taking several low values of $j$ and their explicit $V^{-1}\left[
j\right]  $, Van Kortryk \cite{TSvK}\ was able to deduce the general formula
for the coefficients in the Cayley transform polynomial, which he then
confirmed for specific higher values of $j$.\ \ I provide a proof of the
formula in Section 4 of this paper. \ 

\subsection{Lagrange-Sylvester expansions}

I note in passing another useful method, albeit equivalent to that involving
the inverse Vandermonde matrix. \ 

Reduction of nonsingular matrix functions for spin $j$ to polynomials can be
efficiently carried out in specific cases using Lagrange-Sylvester expansions
\cite{Sylvester}. \ Consider functions of an $N\times N$ diagonalizable matrix
$\mathbb{M}$ with non-degenerate eigenvalues $\lambda_{i}$, $i=1,\cdots,N$.
\ On the span of the eigenvectors, there is an obviously correct Lagrange
formula, as extended to diagonalizable matrices by Sylvester,%
\begin{equation}
f\left(  \mathbb{M}\right)  =\sum_{i=1}^{N}f\left(  \lambda_{i}\right)
\mathbb{P}_{i}\ , \label{39}%
\end{equation}
where the projection operators --- the so-called Frobenius covariants --- are
given by products,%
\begin{equation}
\mathbb{P}_{i}={\prod\limits_{\substack{j=1\\j\neq i}}^{N}}\frac
{\mathbb{M}-\lambda_{j}}{\lambda_{i}-\lambda_{j}}\ . \label{Proj}%
\end{equation}
From expanding each such product it is evident that any $f(\mathbb{M})$
reduces to a polynomial of order $N-1$ in powers of $\mathbb{M}$,
\begin{equation}
f\left(  \mathbb{M}\right)  =\sum_{m=0}^{N-1}C_{m}\left[  f\right]
\ \mathbb{M}^{m}\ , \label{fPoly}%
\end{equation}
and the function-dependent coefficients can be expressed in terms of the
eigenvalues of $\mathbb{M}$\ by expanding the projection operators
(\ref{Proj}) as polynomials in $\mathbb{M}$.

CFZ used the methods described above to obtain the coefficients for the spin
matrix polynomial reduction of the exponential form for rotations for several
low values of $j$. \ They then re-expressed the results in a form that
surprisingly suggested an immediate generalization, after which they developed
the theory to confirm their ansatz, thereby raising it to the level of a
theorem with explicit proofs \cite{CvK}. \ I review and illustrate the CFZ
formula in the next section.\newpage

\section{Rotations as exponentials}

For comparison to the results of the next section, we first recall the CFZ
formula for a rotation through an angle $\theta$ about an axis
$\boldsymbol{\hat{n}}$, valid for any spin $j\in\left\{  0,\frac{1}{2}%
,1,\frac{3}{2},\cdots\right\}  $. \ The formula reduces the \emph{manifestly}
nonsingular exponential function of any spin matrix to an explicit
polynomial:
\begin{equation}
\exp\left(  i~\theta~\boldsymbol{\hat{n}\cdot J}\right)  =\sum_{k=0}^{2j}%
\frac{1}{k!}\left.  A_{k}^{\left[  j\right]  }\left(  \theta\right)  \right.
\left(  2i~\boldsymbol{\hat{n}\cdot J}\right)  ^{k}\ , \label{the result}%
\end{equation}
where the angle-dependent coefficients of the various spin matrix powers are
given by%
\begin{equation}
A_{k}^{\left[  j\right]  }\left(  \theta\right)  =\sin^{k}\left(
\theta/2\right)  ~\left(  \cos\left(  \theta/2\right)  \right)  ^{\epsilon
\left(  j,k\right)  }~\operatorname*{Trunc}_{\left\lfloor j-k/2\right\rfloor
}\left[  \frac{1}{(\sqrt{1-x})^{\epsilon\left(  j,k\right)  }}\left(
\frac{\arcsin\sqrt{x}}{\sqrt{x}}\right)  ^{k}\right]  _{x=\sin^{2}\left(
\theta/2\right)  }\ . \label{exp coefficients}%
\end{equation}
Here, $\left\lfloor \cdots\right\rfloor $ is the integer-valued floor function
and $\operatorname*{Trunc}\limits_{n}\left[  f\left(  x\right)  \right]  $ is
the $n$th-order Taylor polynomial truncation for any $f\left(  x\right)  $
admitting a power series representation:%
\begin{equation}
f\left(  x\right)  =\sum_{m=0}^{\infty}f_{m}x^{m}%
\ ,\ \ \ \operatorname*{Trunc}_{n}\left[  f\left(  x\right)  \right]
\equiv\sum_{m=0}^{n}f_{m}x^{m}\ .
\end{equation}
In addition, $\epsilon\left(  j,k\right)  $ is a binary-valued function of
$2j-k$ that distinguishes even and odd integers: $\ \epsilon\left(
j,k\right)  =0$ for even $2j-k$, and $\epsilon\left(  j,k\right)  =1$ for odd
$2j-k$. \ 

The simplest two nontrivial cases of (\ref{the result}) are well-known:
\ $j=1/2$ and $j=1$. \ Explicitly,%
\begin{align}
\left.  \exp\left(  i~\theta~\boldsymbol{\hat{n}\cdot J}\right)  \right\vert
_{j=1/2}  &  =A_{0}^{\left[  1/2\right]  }\left(  \theta\right)
~\boldsymbol{I}_{2\times2}+2iA_{1}^{\left[  1/2\right]  }\left(
\theta\right)  ~\left(  \boldsymbol{\hat{n}\cdot J}\right)  _{2\times
2}\nonumber\\
&  =\cos\left(  \theta/2\right)  ~\boldsymbol{I}_{2\times2}+2i\sin\left(
\theta/2\right)  ~\left(  \boldsymbol{\hat{n}\cdot J}\right)  _{2\times2}\ ,
\label{ExpSpinHalf}%
\end{align}%
\begin{align}
\left.  \exp\left(  i~\theta~\boldsymbol{\hat{n}\cdot J}\right)  \right\vert
_{j=1}  &  =A_{0}^{\left[  1\right]  }\left(  \theta\right)  ~\boldsymbol{I}%
_{3\times3}+2iA_{1}^{\left[  1\right]  }\left(  \theta\right)  ~\left(
\boldsymbol{\hat{n}\cdot J}\right)  _{3\times3}-2A_{2}^{\left[  1\right]
}\left(  \theta\right)  ~\left(  \boldsymbol{\hat{n}\cdot J}\right)
_{3\times3}^{2}\nonumber\\
&  =\boldsymbol{I}_{3\times3}+i\sin\left(  \theta\right)  \boldsymbol{\left(
\boldsymbol{\hat{n}\cdot J}\right)  }_{3\times3}\boldsymbol{+}\left(
\cos\theta-1\right)  \left(  \boldsymbol{\hat{n}\cdot J}\right)  _{3\times
3}^{2}\ . \label{ExpSpinOne}%
\end{align}
The former involves the Pauli matrices, $\boldsymbol{J}=\boldsymbol{\sigma}%
/2$, while the latter is sometimes known as the Euler-Rodrigues formula.
\ Several other explicit cases may be found in \cite{vanW} and \cite{CFZ}.

In practice, for finite $j$ of reasonable size, the truncations needed to
evaluate (\ref{exp coefficients}) are easily obtained as a matter of course by
machine computation, for example by using either \emph{Maple}%
$^{\textregistered}$ or \emph{Mathematica}$^{\textregistered}$.
\ Nevertheless, it is interesting and useful for a systematic analysis that
Taylor series for powers of cyclometric functions\ can be expressed in terms
of $t\left(  m,n\right)  $, the so-called \emph{central factorial numbers of
the first kind} \cite{Riordan,CFN}. \ Thus for $\left\vert z\right\vert \leq1$
and non-negative integer $n$ (cf. Theorem (4.1.2) in \cite{CFN}), \
\begin{equation}
\left(  \arcsin\left(  z\right)  \right)  ^{n}=\frac{n!}{2^{n}}\sum
_{m=n}^{\infty}\frac{\left\vert t\left(  m,n\right)  \right\vert }{m!}\left(
2z\right)  ^{m}\ . \label{cyclometric power series}%
\end{equation}
Note that the coefficients in these Taylor series are all non-negative. \ In
general, the values of $t\left(  m,n\right)  $\ are defined by and obtained
from simple polynomials, as described in Appendix A.

Incorporating (\ref{cyclometric power series}) into the expression for the
coefficients (\ref{exp coefficients}) gives
\begin{equation}
A_{k}^{\left[  j\right]  }\left(  \theta\right)  =\frac{k!}{2^{k}}\sum
_{m=k}^{2j}\frac{2^{m}}{m!}\left\vert t\left(  m,k\right)  \right\vert
\sin^{m}\left(  \theta/2\right)  \text{ \ \ for even\ }2j-k\ .
\label{the coefficients as cfns}%
\end{equation}
As firmly established in \cite{vanW,CFZ}, the remaining coefficients in
(\ref{the result}) may then be obtained from
\begin{equation}
A_{k-1}^{\left[  j\right]  }\left(  \theta\right)  =\frac{2}{k}\frac
{d}{d\theta}~A_{k}^{\left[  j\right]  }\left(  \theta\right)  \text{ \ \ for
odd\ }2j-k+1\ . \label{oddbydiff}%
\end{equation}

Also, as observed in \cite{CFZ}, the results (\ref{exp coefficients}) display
the limit $j\rightarrow\infty$ for fixed $k$ in a beautifully intuitive way.
\ In that limit, the truncation is lifted to obtain trigonometrical series for
the \emph{periodicized} $\theta^{k}$ monomials. \ But even as $j\rightarrow
\infty$, integer $j$ (bosonic) and semi-integer $j$ (fermionic) coefficients
are clearly distinguished by a relative sign flip for $\theta\in\left[
\pi,3\pi\right]  \operatorname{mod}\left(  4\pi\right)  $. \ This is evident
upon plotting the first few coefficients for very large spins. \ For example,
$A_{0,1,\cdots,5}^{\left[  j\right]  }\left(  \theta\right)  $ are plotted
here for $j=69$ (darker curves, in blue) and $j=137/2$ (lighter curves, in red).

\noindent\hspace{-0.25in}%
{\parbox[b]{3.6588in}{\begin{center}
\includegraphics[
height=2.2707in,
width=3.6588in
]%
{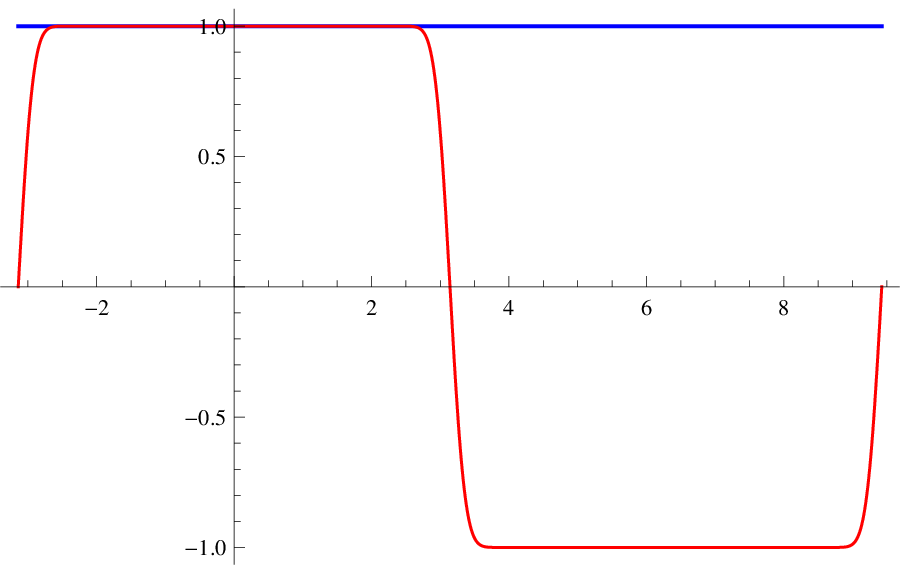}%
\\
$A_{0}^{\left(  j\right)  }\left(  \theta\right)  $ versus $\theta$%
\end{center}}}
\ \ \
{\parbox[b]{3.6588in}{\begin{center}
\includegraphics[
height=2.2707in,
width=3.6588in
]%
{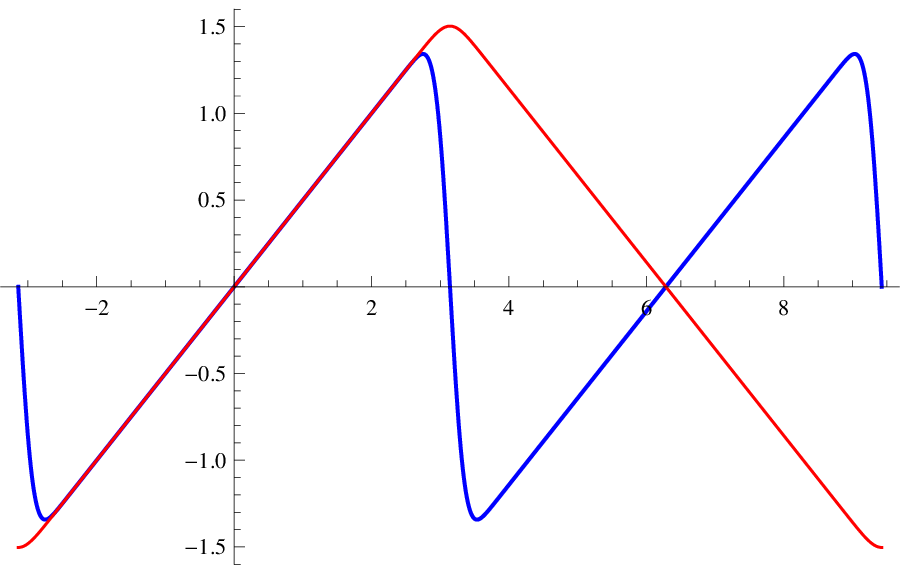}%
\\
$A_{1}^{\left(  j\right)  }\left(  \theta\right)  $ versus $\theta$%
\end{center}}}

\noindent\hspace{-0.25in}%
{\parbox[b]{3.6588in}{\begin{center}
\includegraphics[
height=2.2707in,
width=3.6588in
]%
{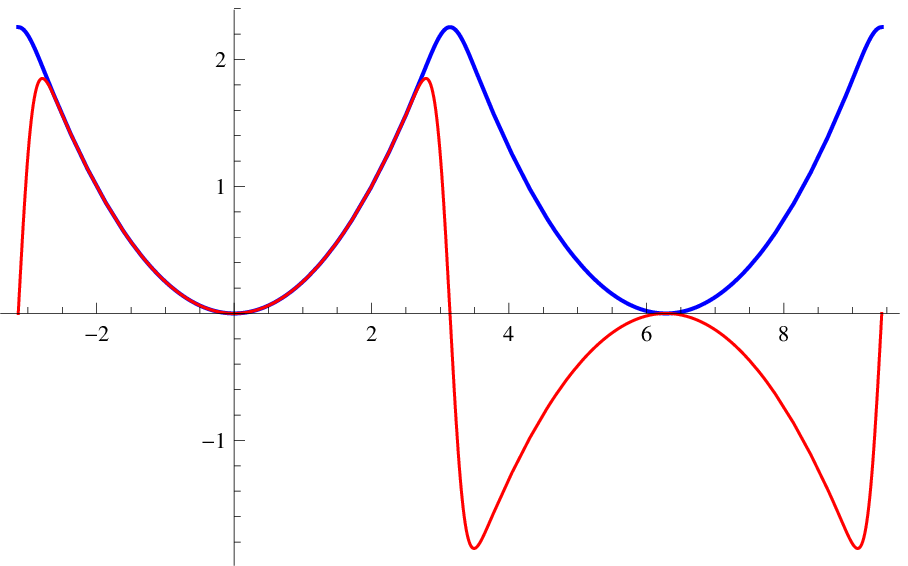}%
\\
$A_{2}^{\left(  j\right)  }\left(  \theta\right)  $ versus $\theta$%
\end{center}}}
\ \ \
{\parbox[b]{3.6588in}{\begin{center}
\includegraphics[
height=2.2707in,
width=3.6588in
]%
{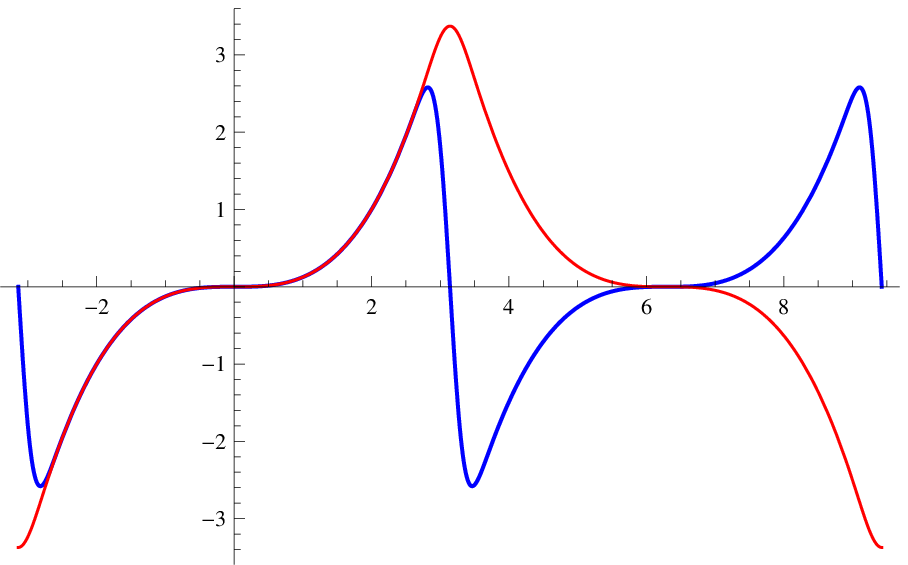}%
\\
$A_{3}^{\left(  j\right)  }\left(  \theta\right)  $ versus $\theta$%
\end{center}}}

\noindent\hspace{-0.25in}%
{\parbox[b]{3.6588in}{\begin{center}
\includegraphics[
height=2.2707in,
width=3.6588in
]%
{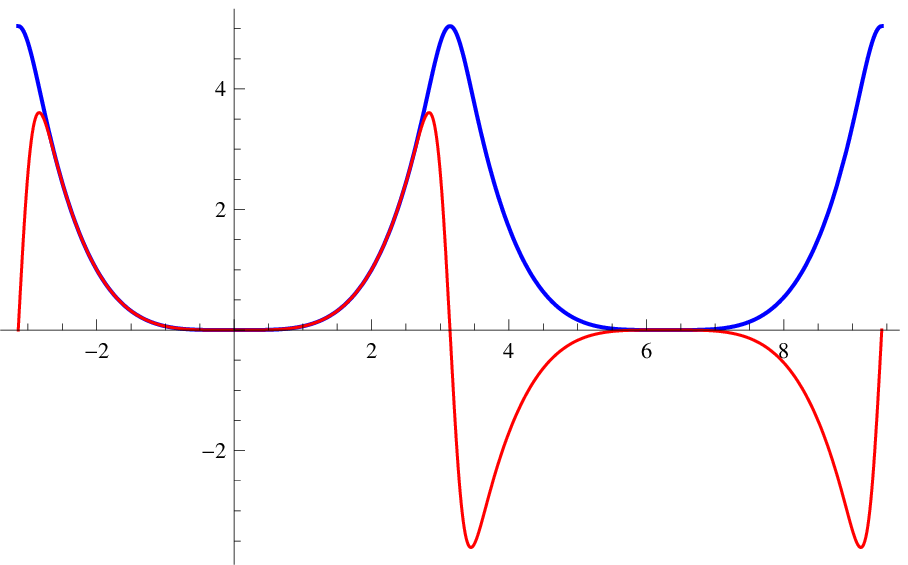}%
\\
$A_{4}^{\left(  j\right)  }\left(  \theta\right)  $ versus $\theta$%
\end{center}}}
\ \ \
{\parbox[b]{3.6588in}{\begin{center}
\includegraphics[
height=2.2707in,
width=3.6588in
]%
{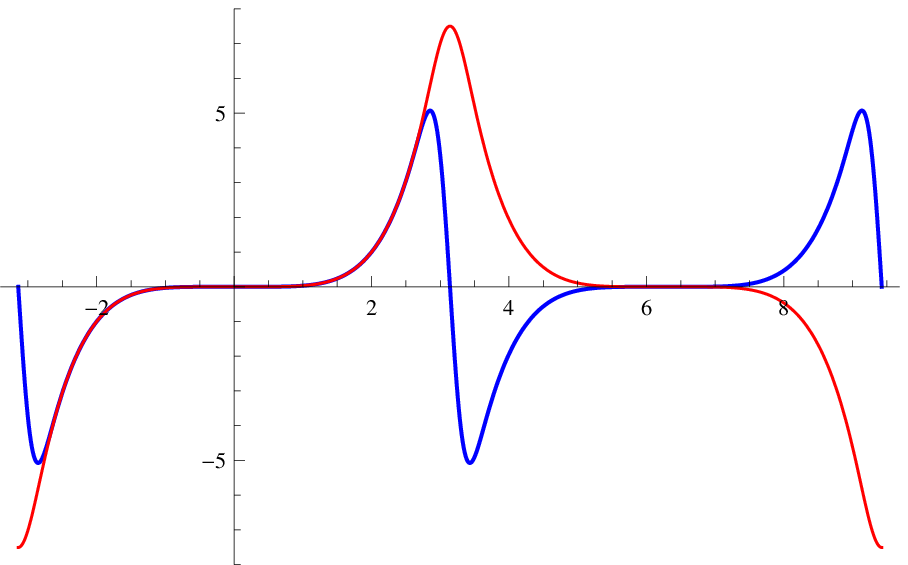}%
\\
$A_{5}^{\left(  j\right)  }\left(  \theta\right)  $ versus $\theta$%
\end{center}}}
\ 

\noindent As stated in (\ref{oddbydiff}), the slope of each red (fermionic)
curve plotted on the right is given exactly by the red curve plotted to its
immediate left. \ Similarly, the slope of any blue (bosonic) curve on the left
is given by the blue curve to its right, but in the row above.

\section{Rotations as Cayley transforms}

The Cayley rational form of a unitary $SU\left(  2\right)  $ group element is
also nonsingular for any irreducible representation with spin $j$. \ Therefore
it can also be reduced to a spin matrix polynomial of order $2j$. \ Thus
\begin{equation}
\frac{1+2i\alpha~\boldsymbol{\hat{n}\cdot J}}{1-2i\alpha~\boldsymbol{\hat
{n}\cdot J}}=\sum_{k=0}^{2j}\left.  \mathfrak{A}_{k}^{\left[  j\right]
}\left(  \alpha\right)  \right.  \left(  2i~\boldsymbol{\hat{n}\cdot
J}\right)  ^{k}\ . \label{Cayley form}%
\end{equation}
Here $\alpha$ is a \emph{real} parameter, to avoid singularities for
\emph{any} $j\in\left\{  0,\frac{1}{2},1,\frac{3}{2},2,\cdots\right\}  $.
\ The coefficients $\mathfrak{A}_{k}^{\left[  j\right]  }\left(
\alpha\right)  $\ are to be determined as functions of $\alpha$. \ These
coefficients can be obtained by rewriting the geometric series $1/\left(
1-2i\alpha~\boldsymbol{\hat{n}\cdot J}\right)  $ for spin $j$ as a polynomial
in $\boldsymbol{\hat{n}\cdot J}$. \ Thus define the auxiliary polynomial%
\begin{equation}
\frac{1}{1-2i\alpha~\boldsymbol{\hat{n}\cdot J}}=\sum_{k=0}^{2j}\left.
\mathfrak{B}_{k}^{\left[  j\right]  }\left(  \alpha\right)  \right.  \left(
2i~\boldsymbol{\hat{n}\cdot J}\right)  ^{k}\ .
\label{GeometricSeriesPolynomial}%
\end{equation}
Then clearly
\begin{equation}
\mathfrak{A}_{0}^{\left[  j\right]  }=2\mathfrak{B}_{0}^{\left[  j\right]
}-1\ ,\ \ \ \mathfrak{A}_{k\geq1}^{\left[  j\right]  }=2\mathfrak{B}_{k\geq
1}^{\left[  j\right]  }\ .
\end{equation}

The simplest two nontrivial cases are again\ $j=1/2$ and $j=1$. \ Explicitly,%
\begin{align}
\left.  \frac{1}{1-2i\alpha~\boldsymbol{\hat{n}\cdot J}}\right\vert _{j=1/2}
&  =\left.  \mathfrak{B}_{0}^{\left[  1/2\right]  }\left(  \alpha\right)
\right.  \boldsymbol{I}_{2\times2}+2i\left.  \mathfrak{B}_{1}^{\left[
1/2\right]  }\left(  \alpha\right)  \right.  \left(  \boldsymbol{\hat{n}\cdot
J}\right)  _{2\times2}\nonumber\\
&  =\frac{1}{1+\alpha^{2}}\left(  \boldsymbol{I}_{2\times2}+2i\alpha~\left(
\boldsymbol{\hat{n}\cdot J}\right)  _{2\times2}\right)  \ ,
\end{align}%
\begin{align}
\left.  \frac{1+2i\alpha~\boldsymbol{\hat{n}\cdot J}}{1-2i\alpha
~\boldsymbol{\hat{n}\cdot J}}\right\vert _{j=1/2}  &  =\left.  \mathfrak{A}%
_{0}^{\left[  1/2\right]  }\left(  \alpha\right)  \right.  \boldsymbol{I}%
_{2\times2}+2i\left.  \mathfrak{A}_{1}^{\left[  1/2\right]  }\left(
\alpha\right)  \right.  \left(  \boldsymbol{\hat{n}\cdot J}\right)
_{2\times2}\nonumber\\
&  =\frac{1}{1+\alpha^{2}}\left(  \left(  1-\alpha^{2}\right)  \boldsymbol{I}%
_{2\times2}+4i\alpha~\left(  \boldsymbol{\hat{n}\cdot J}\right)  _{2\times
2}\right)  \ , \label{CayleySpinHalf}%
\end{align}%
\begin{align}
\left.  \frac{1}{1-2i\alpha~\boldsymbol{\hat{n}\cdot J}}\right\vert _{j=1}  &
=\left.  \mathfrak{B}_{0}^{\left[  1\right]  }\left(  \alpha\right)  \right.
\boldsymbol{I}_{3\times3}+2i\left.  \mathfrak{B}_{1}^{\left[  1\right]
}\left(  \alpha\right)  \right.  \boldsymbol{\left(  \boldsymbol{\hat{n}\cdot
J}\right)  }_{3\times3}-4\left.  \mathfrak{B}_{2}^{\left[  1\right]  }\left(
\alpha\right)  \right.  \left(  \boldsymbol{\hat{n}\cdot J}\right)
_{3\times3}^{2}\nonumber\\
&  =\frac{1}{1+4\alpha^{2}}\left(  \left(  1+4\alpha^{2}\right)
\boldsymbol{I}_{3\times3}+2i\alpha~\boldsymbol{\left(  \boldsymbol{\hat
{n}\cdot J}\right)  }_{3\times3}\boldsymbol{-}4\alpha^{2}\left(
\boldsymbol{\hat{n}\cdot J}\right)  _{3\times3}^{2}\right)  \ ,
\end{align}%
\begin{align}
\left.  \frac{1+2i\alpha~\boldsymbol{\hat{n}\cdot J}}{1-2i\alpha
~\boldsymbol{\hat{n}\cdot J}}\right\vert _{j=1}  &  =\left.  \mathfrak{A}%
_{0}^{\left[  1\right]  }\left(  \alpha\right)  \right.  \boldsymbol{I}%
_{3\times3}+2i\left.  \mathfrak{A}_{1}^{\left[  1\right]  }\left(
\alpha\right)  \right.  \boldsymbol{\left(  \boldsymbol{\hat{n}\cdot
J}\right)  }_{3\times3}-4\left.  \mathfrak{A}_{2}^{\left[  1\right]  }\left(
\alpha\right)  \right.  \left(  \boldsymbol{\hat{n}\cdot J}\right)
_{3\times3}^{2}\nonumber\\
&  =\frac{1}{1+4\alpha^{2}}\left(  \left(  1+4\alpha^{2}\right)
\boldsymbol{I}_{3\times3}+4i\alpha~\boldsymbol{\left(  \boldsymbol{\hat
{n}\cdot J}\right)  }_{3\times3}\boldsymbol{-}8\alpha^{2}\left(
\boldsymbol{\hat{n}\cdot J}\right)  _{3\times3}^{2}\right)  \ .
\label{CayleySpinOne}%
\end{align}
Yet again, the first two of these polynomials involve the Pauli matrices,
$\boldsymbol{J}=\boldsymbol{\sigma}/2$, while the last is related to the
Cayley transform of the Euler-Rodrigues formula. \ Appendix D repeats these
two examples and also gives the Cayley transforms for $j=3/2,\ 2,\ 5/2,$ and
$3$.

The coefficients in the geometric series expansion for arbitrary integer or
semi-integer $j$ follow directly from the methods in \cite{CFZ,CvK}. \ The
result is \cite{TSvK}%
\begin{equation}
\mathfrak{B}_{k}^{\left[  j\right]  }\left(  \alpha\right)  =\frac{\alpha^{k}%
}{\det\left(  1-2i\alpha~\boldsymbol{\hat{n}\cdot J}\right)  }%
~\operatorname*{Trunc}_{2j-k}\left[  \det\left(  1-2i\alpha~\boldsymbol{\hat
{n}\cdot J}\right)  \right]  \ , \label{GSPCoefficients}%
\end{equation}
where the truncation is in powers of $\alpha$, and where the determinant for
spin $j$ is%
\begin{equation}
\det\left(  1-2i\alpha~\boldsymbol{\hat{n}\cdot J}\right)  =%
{\displaystyle\prod\limits_{m=0}^{2j}}
\left(  1-2i\alpha\left(  j-m\right)  \right)  =%
{\displaystyle\prod\limits_{n=1}^{\left\lfloor j+1/2\right\rfloor }}
\left(  1+4\alpha^{2}\left(  j+1-n\right)  ^{2}\right)  \ .
\label{Determinant}%
\end{equation}
These results are readily checked for small values of $j$ upon using explicit
matrices, say $\boldsymbol{\hat{n}\cdot J}=J_{3}$. \ I provide a detailed
proof of (\ref{GSPCoefficients})\ in the following subsection.

But first, note for integer spins it is always true that $\mathfrak{B}%
_{0}^{\left[  j\right]  }\left(  \alpha\right)  =1$, and that the higher
coefficients are paired, \
\begin{equation}
\mathfrak{B}_{2k+2}^{\left[  j\right]  }\left(  \alpha\right)  =\alpha
~\mathfrak{B}_{2k+1}^{\left[  j\right]  }\left(  \alpha\right)  \text{ \ \ for
\ \ }0\leq k\leq j-1\ \ \ \text{and\ integer }j\ , \label{BosonicPairing}%
\end{equation}
while \emph{all} the coefficients are paired for semi-integer spins. \ In this
case%
\begin{equation}
\mathfrak{B}_{2k+1}^{\left[  j\right]  }\left(  \alpha\right)  =\alpha
~\mathfrak{B}_{2k}^{\left[  j\right]  }\left(  \alpha\right)  \text{ \ \ for
\ \ }0\leq k\leq j-\frac{1}{2}\ \ \ \text{and\ semi-integer }j\ .
\label{FermionicPairing}%
\end{equation}
So, the coefficients in the spin matrix polynomials for bosonic (integer $j$)
and fermionic (semi-integer $j$) Cayley transforms are easily distinguished,
both quantitatively and qualitatively. \ The coefficients are paired as
indicated because, for either bosonic or fermionic spins, only \emph{even}
powers of $\alpha$ are produced by the determinant factors in
(\ref{GSPCoefficients}). \ Hence the parities: $\ \mathfrak{A}_{k}^{\left[
j\right]  }\left(  -\alpha\right)  =\left(  -1\right)  ^{k}\mathfrak{A}%
_{k}^{\left[  j\right]  }\left(  \alpha\right)  $ and $\mathfrak{B}%
_{k}^{\left[  j\right]  }\left(  -\alpha\right)  =\left(  -1\right)
^{k}\mathfrak{B}_{k}^{\left[  j\right]  }\left(  \alpha\right)  $.

Moreover, for any allowed $j$\ all the coefficients of $\alpha^{2k}$ in
(\ref{Determinant}) are positive. \ As a consequence, for any allowed $j$ the
$\mathfrak{A}_{k}^{\left[  j\right]  }\left(  \alpha\right)  $ and
$\mathfrak{B}_{k}^{\left[  j\right]  }\left(  \alpha\right)  $\ coefficients
have no singularities for real $\alpha$. \ In particular, for any $j$ the
highest two coefficients reduce to
\begin{equation}
\frac{1}{\alpha^{2j}}\left.  \mathfrak{B}_{2j}^{\left[  j\right]  }\left(
\alpha\right)  \right.  =\frac{1}{\alpha^{2j-1}}\left.  \mathfrak{B}%
_{2j-1}^{\left[  j\right]  }\left(  \alpha\right)  \right.  \mathfrak{=}%
\frac{1}{\det\left(  1-2i\alpha~\boldsymbol{\hat{n}\cdot J}\right)  }\text{ .}%
\end{equation}
And as $j$ increases, the domain where $1/\det\left(  1-2i\alpha
~\boldsymbol{\hat{n}\cdot J}\right)  $ achieves significant values collapses
towards the origin, at which point it always has unit value. \ Here is a plot
for $j=1/2,\ 1,\ 3/2,\ 2,$ $5/2,$ \& $3$.%
\begin{center}
\includegraphics[
height=3.1506in,
width=4.7223in
]%
{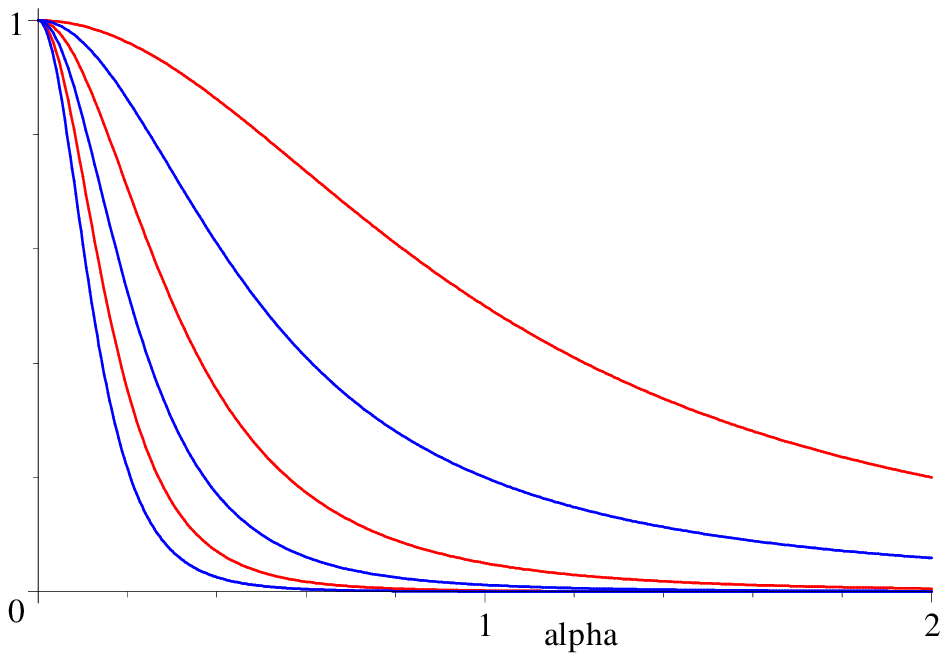}%
\\
$1/\det\left(  1-2i\alpha~\boldsymbol{\hat{n}\cdot J}\right)  $ plotted for
$j=\frac{1}{2},1,\frac{3}{2},2,$ $\frac{5}{2}$, and $3,$ for upper to lower
curves, respectively.
\end{center}

Two additional comments are warranted \cite{TSvK}. \ First, the determinants
in (\ref{GSPCoefficients}) are essentially generating functions of the central
factorial numbers $t\left(  m,n\right)  $ (see Appendix A), a fact previously
exploited in \cite{CFZ,CvK} and already emphasized for the Cayley transforms
in \cite{TSvK}. \ For either integer or semi-integer $j$,%
\begin{equation}
\operatorname*{Trunc}_{2j-k}\left[  \det\left(  1-2i\alpha~\boldsymbol{\hat
{n}\cdot J}\right)  \right]  =\sum_{m=0}^{\left\lfloor j-k/2\right\rfloor
}4^{m}\alpha^{2m}~\left\vert t\left(  2j+2,2j+2-2m\right)  \right\vert \ ,
\label{TruncsAsCFNs}%
\end{equation}
with the full determinant obtained for $k=0$ if $j$ is integer. \ For
semi-integer $j$, the same expression is true for truncations, only now the
full determinant includes an additional term $4^{j+\frac{1}{2}}\alpha
^{2j+1}~\left\vert t\left(  2j+2,1\right)  \right\vert $. \ Thus the general
formula for the determinant, for either integer or semi-integer $j$, is
\begin{equation}
\det\left(  1-2i\alpha~\boldsymbol{\hat{n}\cdot J}\right)  =\sum
_{k=0}^{\left\lfloor j+\frac{1}{2}\right\rfloor }4^{k}\alpha^{2k}~\left\vert
t\left(  2j+2,2j+2-2k\right)  \right\vert \ . \label{DeterminantAsCFNs}%
\end{equation}
Second, as $j\rightarrow\infty$ for any fixed $k$ the truncation in
(\ref{GSPCoefficients}) is lifted --- but with some subtleties to be discussed
below ---\ to obtain for small $\alpha$, $\lim_{j\rightarrow\infty
}\mathfrak{B}_{k}^{\left[  j\right]  }\left(  \alpha\right)  \sim\alpha^{k}$.
\ But in contrast to the periodicized $\theta$-monomials found in \cite{CFZ},
the large $j$ behavior here does not make the periodicity of rotations
manifest. \ To exhibit periodicity even for finite values of $j$, $\theta$
must be expressed, on a case-by-case basis, as cyclometric functions of
$\alpha$, and then the branch structure of those cyclometric functions must be
invoked \cite{TSvK}.

\subsection{A derivation for general matrix resolvents}

In this subsection I obtain the coefficients in the Cayley transform spin
matrix polynomial by deriving the expansion coefficients for the resolvent of
a general matrix. \ This is a well-known result in resolvent theory (e.g. see
\cite{He}) and has been established many times in the literature (e.g. see
\cite{Householder,Faddeev}). \ While there is hardly any need to plow again a
field that has been tilled as much as this one, I include here a variant of
the standard proof just to make the discussion self-contained. \ (Also see
Appendix D.)

For any finite $N\times N$\ matrix $M$ consider the resolvent written as a
matrix polynomial,%
\begin{equation}
\frac{1}{1-\alpha M}=\sum_{m=0}^{N-1}r_{m}\left(  \alpha\right)  ~M^{m}\ .
\label{ResolventMatrixPoly}%
\end{equation}
This is much simpler than the exponential case \cite{CFZ,CvK} in that one
\emph{need not} differentiate, e.g. to find%
\begin{equation}
\left(  1-\alpha M\right)  \frac{d}{d\alpha}\left(  \frac{1}{1-\alpha
M}\right)  =\frac{M}{1-\alpha M}\ ,
\end{equation}
so as to obtain an expression linear in $M$ multiplying the original
polynomial, and subsequently a set of first-order equations for the
coefficients $r_{m}$. \ Rather, it suffices here just to multiply through by
$1-\alpha M$. \ Thus
\begin{equation}
1=\left(  1-\alpha M\right)  \frac{1}{1-\alpha M}=\left(  1-\alpha M\right)
\sum_{m=0}^{N-1}M^{m}r_{m}\ . \label{Only1}%
\end{equation}
On the other hand, the Cayley-Hamilton theorem gives%
\begin{equation}
\sum_{m=0}^{N}M^{m}d_{m}=0\ ,\ \ \ M^{N}=-\frac{1}{d_{N}}\sum_{m=0}^{N-1}%
M^{m}d_{m}\ ,
\end{equation}
where the polynomial coefficients $d_{n}$ are defined by the determinant
\begin{equation}
\det\left(  1-\alpha M\right)  =\sum_{m=0}^{N}\alpha^{m}d_{m}\ ,
\end{equation}
with $d_{0}=1$, $d_{N}=\left(  -1\right)  ^{N}\det M$, etc. (see
\cite{GalileonPrimer}). \ Therefore (\ref{Only1}) becomes%
\begin{equation}
1=r_{0}+\frac{\alpha r_{N-1}}{d_{N}}~d_{0}+\sum_{m=1}^{N-1}M^{m}\left(
r_{m}-\alpha r_{m-1}+\frac{\alpha r_{N-1}}{d_{N}}~d_{m}\right)  \ .
\end{equation}
If the powers of $M^{m}$ for $0\leq m\leq N-1$ are all linearly independent,
as is the case for the spin matrices $\boldsymbol{\hat{n}\cdot J}$, then
\begin{equation}
r_{0}+\frac{\alpha r_{N-1}}{d_{N}}~d_{0}=1\text{ \ \ and}\ \ \ r_{m}-\alpha
r_{m-1}+\frac{\alpha r_{N-1}}{d_{N}}~d_{m}=0\text{ \ \ for }1\leq m\leq N-1\ .
\end{equation}
It is now straightforward to show the solution to these recursion relations is
given by
\begin{equation}
r_{n}=\frac{\alpha^{n}}{\det\left(  1-\alpha M\right)  }~\operatorname*{Trunc}%
_{N-1-n}\left[  \det\left(  1-\alpha M\right)  \right]  \ .
\end{equation}
For $M=2i\boldsymbol{\hat{n}\cdot J}$ with $N=2j+1$, the $r_{n}$ are
essentially just renamed $\mathfrak{B}_{n}^{\left[  j\right]  }$. \ Thus the
result (\ref{GSPCoefficients})\ is proven.

\subsection{Additional exact results and asymptotic behavior for large $j$}

The determinants can also be written as Pochhammer symbols: $\ x^{\left(
n\right)  }\equiv x\left(  x+1\right)  \cdots\left(  x+n-1\right)
=\frac{\Gamma\left(  x+n\right)  }{\Gamma\left(  x\right)  }$. \ For example,
for integer $j$,%
\begin{align}%
{\displaystyle\prod\limits_{m=0}^{j}}
\left(  1-2i\alpha\left(  j-m\right)  \right)   &  =%
{\displaystyle\prod\limits_{n=1}^{j}}
\left(  1-2i\alpha n\right)  =\left(  -2i\alpha\right)  ^{j+1}\frac
{\Gamma\left(  j+1-\frac{1}{2i\alpha}\right)  }{\Gamma\left(  -\frac
{1}{2i\alpha}\right)  }\ ,\\%
{\displaystyle\prod\limits_{m=j+1}^{2j}}
\left(  1-2i\alpha\left(  j-m\right)  \right)   &  =%
{\displaystyle\prod\limits_{n=1}^{j}}
\left(  1+2i\alpha n\right)  =\left(  2i\alpha\right)  ^{j+1}\frac
{\Gamma\left(  j+1+\frac{1}{2i\alpha}\right)  }{\Gamma\left(  \frac
{1}{2i\alpha}\right)  }\ ,
\end{align}
which leads to a nice form for the determinant expressed in terms of analytic
functions of both $j$ and $\alpha$. \ Again, for integer $j$,%
\begin{align}
\det\left(  1-2i\alpha~\boldsymbol{\hat{n}\cdot J}\right)   &  =\left(
4\alpha^{2}\right)  ^{j+1}\frac{\Gamma\left(  j+1-\frac{1}{2i\alpha}\right)
}{\Gamma\left(  -\frac{1}{2i\alpha}\right)  }\frac{\Gamma\left(  j+1+\frac
{1}{2i\alpha}\right)  }{\Gamma\left(  \frac{1}{2i\alpha}\right)  }\nonumber\\
&  =\frac{1}{\pi}\left(  2\alpha\right)  ^{2j+1}\sinh\left(  \frac{\pi
}{2\alpha}\right)  \times\left\vert \Gamma\left(  j+1+\frac{1}{2i\alpha
}\right)  \right\vert ^{2}\ , \label{BosonicClosedFormDeterminant}%
\end{align}
upon using $\Gamma\left(  \frac{1}{2i\alpha}\right)  \Gamma\left(  -\frac
{1}{2i\alpha}\right)  =2\pi\alpha/\sinh\frac{\pi}{2\alpha}$. \ On the other
hand, for semi-integer $j$,%
\begin{equation}
\det\left(  1-2i\alpha~\boldsymbol{\hat{n}\cdot J}\right)  =\frac{1}{\pi
}\left(  2\alpha\right)  ^{2j+1}\cosh\left(  \frac{\pi}{2\alpha}\right)
\left\vert \Gamma\left(  j+1+\frac{1}{2i\alpha}\right)  \right\vert ^{2}\ .
\label{FermionicClosedFormDeterminant}%
\end{equation}
In the limit $j\rightarrow\infty$, these are the well-known infinite product
representations of $\sinh$ and $\cosh$. \ That is to say,%
\begin{equation}
\lim_{j\rightarrow\infty}\frac{\det\left(  1-2i\alpha~\boldsymbol{\hat{n}\cdot
J}\right)  }{\left(  4\alpha^{2}\right)  ^{\left\lfloor j+\frac{1}%
{2}\right\rfloor }\left(  \Gamma\left(  1+j\right)  \right)  ^{2}}=\left\{
\begin{array}
[c]{ccc}%
\frac{2\alpha}{\pi}\sinh\left(  \frac{\pi}{2\alpha}\right)  &  & \text{for
integer\ }j\\
&  & \\
\frac{1}{\pi}\cosh\left(  \frac{\pi}{2\alpha}\right)  &  & \text{for
semi-integer\ }j
\end{array}
\right.  \ . \label{AsymptoticClosedFormDeterminants}%
\end{equation}
This shows that there is a distinguishable difference between Cayley
transforms for bosonic and fermionic spins, even as $j\rightarrow\infty$, much
as there were distinguishable differences in the behavior of the coefficients
for the exponential (\ref{the result}), also as $j\rightarrow\infty$, as
illustrated in the Figures of Section 3.

Now consider the truncations for $k=1$ or $2$, say for integer spin. \ This
choice of coefficients removes from $\det\left(  1-2i\alpha~\boldsymbol{\hat
{n}\cdot J}\right)  $ the highest power of $\alpha$, namely the $\alpha^{2j}$
term. \ But this term is easily computed:%
\begin{equation}%
{\displaystyle\prod\limits_{n=1}^{j}}
\left(  4\alpha^{2}\left(  j+1-n\right)  ^{2}\right)  =%
{\displaystyle\prod\limits_{k=1}^{j}}
\left(  4\alpha^{2}k^{2}\right)  =4^{j}\alpha^{2j}\left(  j!\right)
^{2}=4^{j}\alpha^{2j}\left\vert \Gamma\left(  j+1\right)  \right\vert ^{2}\ .
\end{equation}
Therefore for integer $j$%
\begin{align}
\operatorname*{Trunc}_{2j-1}\left[  \det\left(  1-2i\alpha~\boldsymbol{\hat
{n}\cdot J}\right)  \right]   &  =\operatorname*{Trunc}_{2j-2}\left[
\det\left(  1-2i\alpha~\boldsymbol{\hat{n}\cdot J}\right)  \right]
=\det\left(  1-2i\alpha~\boldsymbol{\hat{n}\cdot J}\right)  -4^{j}\alpha
^{2j}\left(  j!\right)  ^{2}\nonumber\\
&  =4^{j}\alpha^{2j}\left(  \frac{2\alpha}{\pi}\sinh\left(  \frac{\pi}%
{2\alpha}\right)  \times\left\vert \Gamma\left(  j+1+\frac{1}{2i\alpha
}\right)  \right\vert ^{2}-\left\vert \Gamma\left(  j+1\right)  \right\vert
^{2}\right)  \ .
\end{align}
The corresponding coefficients in the Cayley rational form are%
\begin{align}
\mathfrak{B}_{1}^{\left[  j\right]  }\left(  \alpha\right)   &  =\frac{\alpha
}{\det\left(  1-2i\alpha~\boldsymbol{\hat{n}\cdot J}\right)  }\left(
\det\left(  1-2i\alpha~\boldsymbol{\hat{n}\cdot J}\right)  -4^{j}\alpha
^{2j}\left(  j!\right)  ^{2}\right) \nonumber\\
&  =\alpha\left(  1-\frac{4^{j}\alpha^{2j}\left\vert \Gamma\left(  j+1\right)
\right\vert ^{2}}{\frac{1}{\pi}\left(  2\alpha\right)  ^{2j+1}\sinh\left(
\frac{\pi}{2\alpha}\right)  \times\left\vert \Gamma\left(  j+1+\frac
{1}{2i\alpha}\right)  \right\vert ^{2}}\right)  \ .
\end{align}
Similarly for $\mathfrak{B}_{2}^{\left[  j\right]  }\left(  \alpha\right)
=\alpha~\mathfrak{B}_{1}^{\left[  j\right]  }\left(  \alpha\right)  $. \ The
exact results for these two coefficients, for all integer spins, are then
given by%
\begin{equation}
\frac{1}{\alpha}\left.  \mathfrak{B}_{1}^{\left[  j\right]  }\left(
\alpha\right)  \right.  =\frac{1}{\alpha^{2}}\left.  \mathfrak{B}_{2}^{\left[
j\right]  }\left(  \alpha\right)  \right.  =1-\frac{\pi~\left\vert
\Gamma\left(  j+1\right)  \right\vert ^{2}}{2\alpha\sinh\left(  \frac{\pi
}{2\alpha}\right)  \times\left\vert \Gamma\left(  j+1+\frac{1}{2i\alpha
}\right)  \right\vert ^{2}}\ .
\end{equation}
For fixed $\alpha$, the large $j$ limits of these follow immediately from
\begin{equation}
\lim_{j\rightarrow\infty}\frac{\left\vert \Gamma\left(  j+1\right)
\right\vert ^{2}}{\left\vert \Gamma\left(  j+1+\frac{1}{2i\alpha}\right)
\right\vert ^{2}}=1\ .
\end{equation}
In fact, the limit is just $1$ to all orders in $\frac{1}{\alpha}$. \ This
should be obvious, but in case it is not, as a check the reader may consider
the following expansion:
\begin{align}
\frac{\left(  \Gamma\left(  j+1\right)  \right)  ^{2}}{\Gamma\left(
j+1+z\right)  \Gamma\left(  j+1-z\right)  }  &  =1-\operatorname{Psi}\left(
1,j+1\right)  z^{2}+\frac{1}{12}\left(  6\left(  \operatorname{Psi}\left(
1,j+1\right)  \right)  ^{2}-\operatorname{Psi}\left(  3,j+1\right)  \right)
z^{4}\\
&  -\frac{1}{360}\left(  \operatorname{Psi}\left(  5,j+1\right)
-30\operatorname{Psi}\left(  3,j+1\right)  \operatorname{Psi}\left(
1,j+1\right)  +60\left(  \operatorname{Psi}\left(  1,j+1\right)  \right)
^{3}\right)  z^{6}\nonumber\\
&  +\frac{1}{20\,160}\left(
\begin{array}
[c]{c}%
70\left(  \operatorname{Psi}\left(  3,j+1\right)  \right)  ^{2}+840\left(
\operatorname{Psi}\left(  1,j+1\right)  \right)  ^{4}-\operatorname{Psi}%
\left(  7,j+1\right) \\
+56\operatorname{Psi}\left(  5,j+1\right)  \operatorname{Psi}\left(
1,j+1\right)  -840\operatorname{Psi}\left(  3,j+1\right)  \left(
\operatorname{Psi}\left(  1,j+1\right)  \right)  ^{2}%
\end{array}
\right)  z^{8}\nonumber\\
&  +O\left(  z^{10}\right)  \ ,\nonumber
\end{align}
where $\operatorname{Psi}\left(  n,z\right)  $ is the $n$th derivative of the
digamma function $\operatorname{Psi}\left(  z\right)  $ (also known as
\textsf{PolyGamma} in \emph{Mathematica}$^{\textregistered}$). \ Except for
the $O\left(  z^{0}\right)  $ term, all of the $O\left(  z^{k}\right)  $
coefficients in this last expansion vanish as $j\rightarrow\infty$. \ 

Therefore it follows for integer $j$ that%
\begin{equation}
\frac{1}{\alpha}\left.  \mathfrak{B}_{1}^{\left[  j\right]  }\left(
\alpha\right)  \right.  =\frac{1}{\alpha^{2}}\left.  \mathfrak{B}_{2}^{\left[
j\right]  }\left(  \alpha\right)  \right.  \underset{j\rightarrow
\infty}{\longrightarrow}1-\frac{\pi}{2\alpha\sinh\left(  \frac{\pi}{2\alpha
}\right)  }\ .
\end{equation}
So, as $\alpha$ approaches zero, the RHS approaches unity faster than any
power of $\alpha$ due to the essential singularity in the exponential,
\begin{equation}
\frac{\pi}{2\alpha\sinh\left(  \frac{\pi}{2\alpha}\right)  }\underset{\alpha
\rightarrow0}{\sim}\frac{\pi}{\left\vert \alpha\right\vert }~e^{-\frac{\pi
}{2\left\vert \alpha\right\vert }}\ .
\end{equation}
This is the precise meaning of the statement in \cite{TSvK} that%
\begin{equation}
\lim_{j\rightarrow\infty}\mathfrak{B}_{1,2}^{\left[  j\right]  }\left(
\alpha\right)  \underset{\alpha\rightarrow0}{\sim}\alpha^{1,2}\ .
\end{equation}
In fact, the asymptotic form is rapidly approached as $j$ is increased. \ 

A similar calculation, again for integer $j$, gives%
\begin{align}
\frac{1}{\alpha^{3}}\left.  \mathfrak{B}_{3}^{\left[  j\right]  }\left(
\alpha\right)  \right.   &  =\frac{1}{\alpha^{4}}\left.  \mathfrak{B}%
_{4}^{\left[  j\right]  }\left(  \alpha\right)  \right.  =1-\frac
{\pi~\left\vert \Gamma\left(  j+1\right)  \right\vert ^{2}}{2\alpha
\sinh\left(  \frac{\pi}{2\alpha}\right)  \times\left\vert \Gamma\left(
j+1+\frac{1}{2i\alpha}\right)  \right\vert ^{2}}\left(  1+\frac{1}%
{24\alpha^{2}}\left(  \pi^{2}-6\operatorname{Psi}\left(  1,1+j\right)
\right)  \right) \\
&  \underset{j\rightarrow\infty}{\longrightarrow}1-\frac{\pi}{2\alpha
\sinh\left(  \frac{\pi}{2\alpha}\right)  }\left(  1+\frac{\pi^{2}}%
{24\alpha^{2}}\right)  \ .
\end{align}
\ For the case at hand,%
\begin{equation}
\operatorname{Psi}\left(  1,1+j\right)  =\zeta\left(  2\right)  -\sum
_{k=1}^{j}\frac{1}{k^{2}}\ .
\end{equation}

Here are some other facts that were used to get exact expressions as well as
asymptotic behaviors for the $\mathfrak{B}_{1-4}^{\left[  j\right]  }\left(
\alpha\right)  $ coefficients:%
\begin{equation}
\prod\limits_{l=0}^{j}\left(  z+l^{2}\right)  =\frac{\Gamma\left(
j+1-\sqrt{-z}\right)  \Gamma\left(  j+1+\sqrt{-z}\right)  }{\Gamma\left(
-\sqrt{-z}\right)  \Gamma\left(  \sqrt{-z}\right)  }\ ,
\end{equation}%
\begin{equation}
\left\vert t\left(  2j+2,2\right)  \right\vert =\left.  \frac{d}{dz}%
\prod\limits_{l=0}^{j}\left(  z+l^{2}\right)  \right\vert _{z=0}=\left(
\Gamma\left(  j+1\right)  \right)  ^{2}\ ,
\end{equation}%
\begin{equation}
\left\vert t\left(  2j+2,4\right)  \right\vert =\frac{1}{2}\left.  \frac
{d^{2}}{dz^{2}}\prod\limits_{l=0}^{j}\left(  z+l^{2}\right)  \right\vert
_{z=0}=\frac{1}{6}\left(  \Gamma\left(  j+1\right)  \right)  ^{2}\pi
^{2}-\left(  \Gamma\left(  j+1\right)  \right)  ^{2}\operatorname{Psi}\left(
1,1+j\right)  \ ,
\end{equation}%
\begin{gather}
\frac{\operatorname*{Trunc}_{2j-1}\left[  \det\left(  1-2ix~\boldsymbol{\hat
{n}\cdot J}\right)  \right]  }{\det\left(  1-2ix~\boldsymbol{\hat{n}\cdot
J}\right)  }=\frac{\operatorname*{Trunc}_{2j-2}\left[  \det\left(
1-2ix~\boldsymbol{\hat{n}\cdot J}\right)  \right]  }{\det\left(
1-2ix~\boldsymbol{\hat{n}\cdot J}\right)  }=1-\frac{4^{j}x^{2j}\left(
\Gamma\left(  j+1\right)  \right)  ^{2}}{\det\left(  1-2ix~\boldsymbol{\hat
{n}\cdot J}\right)  }\nonumber\\
=1-\frac{\left(  \Gamma\left(  j+1\right)  \right)  ^{2}}{\frac{2x}{\pi
}\left(  \sinh\frac{\pi}{2x}\right)  \left\vert \Gamma\left(  j+1+\frac{i}%
{2x}\right)  \right\vert ^{2}}\ ,
\end{gather}%
\begin{gather}
\frac{\operatorname*{Trunc}_{2j-3}\left[  \det\left(  1-2ix~\boldsymbol{\hat
{n}\cdot J}\right)  \right]  }{\det\left(  1-2ix~\boldsymbol{\hat{n}\cdot
J}\right)  }=\frac{\operatorname*{Trunc}_{2j-4}\left[  \det\left(
1-2ix~\boldsymbol{\hat{n}\cdot J}\right)  \right]  }{\det\left(
1-2ix~\boldsymbol{\hat{n}\cdot J}\right)  }=1-\frac{4^{j}x^{2j}\left(
\Gamma\left(  j+1\right)  \right)  ^{2}}{\det\left(  1-2ix~\boldsymbol{\hat
{n}\cdot J}\right)  }\nonumber\\
-\frac{4^{j-1}x^{2j-2}\left(  \frac{1}{6}\left(  \Gamma\left(  j+1\right)
\right)  ^{2}\pi^{2}-\left(  \Gamma\left(  j+1\right)  \right)  ^{2}%
\operatorname{Psi}\left(  1,1+j\right)  \right)  }{\det\left(
1-2ix~\boldsymbol{\hat{n}\cdot J}\right)  }\nonumber\\
=1-\frac{\pi~\left\vert \Gamma\left(  j+1\right)  \right\vert ^{2}}%
{2\alpha\sinh\left(  \frac{\pi}{2\alpha}\right)  \times\left\vert
\Gamma\left(  j+1+\frac{1}{2i\alpha}\right)  \right\vert ^{2}}\left(
1+\frac{1}{24\alpha^{2}}\left(  \pi^{2}-6\operatorname{Psi}\left(
1,1+j\right)  \right)  \right)  \ .
\end{gather}
\newpage

I plot a few examples to illustrate these features, beginning with
$\mathfrak{B}_{1}^{\left[  j\right]  }\left(  \alpha\right)  $\ and
$\mathfrak{B}_{2}^{\left[  j\right]  }\left(  \alpha\right)  $.%
\begin{center}
\includegraphics[
height=3.5994in,
width=5.4106in
]%
{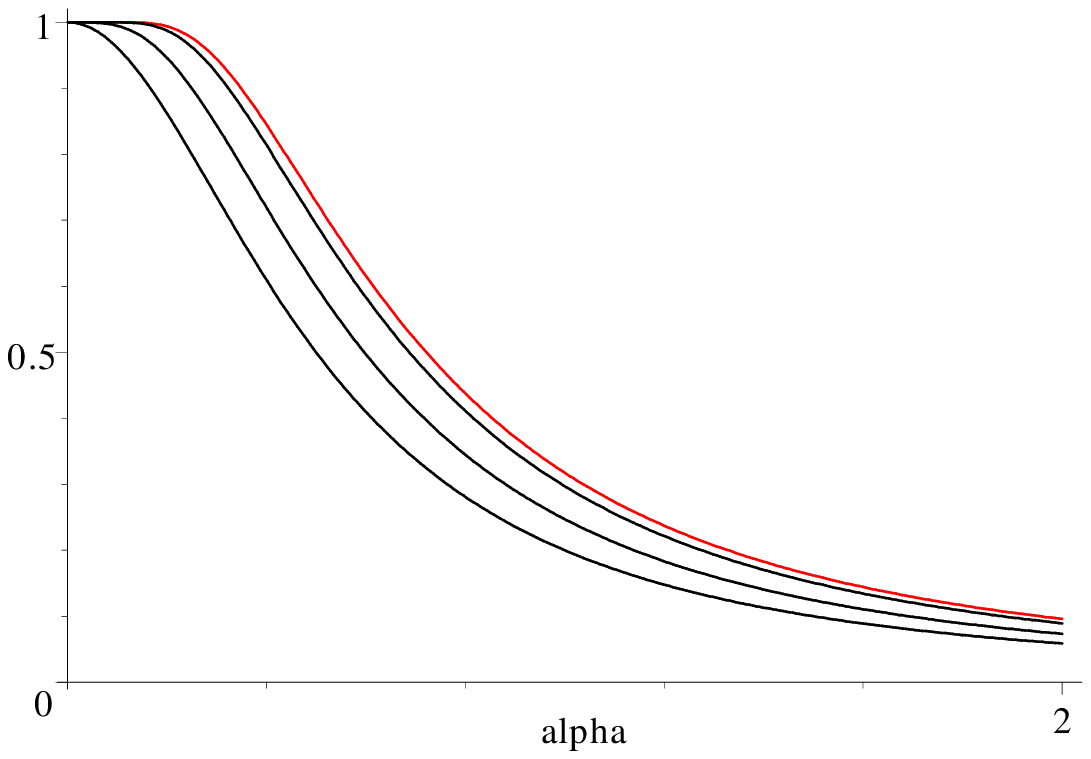}%
\\
$\left.  \mathfrak{B}_{1}^{\left[  j\right]  }\left(  \alpha\right)  \right.
/\alpha=\left.  \mathfrak{B}_{2}^{\left[  j\right]  }\left(  \alpha\right)
\right.  /\alpha^{2}$ plotted for $j=1,2,$ \& $8$ (lower to upper black
curves) and the $j\rightarrow\infty$ limit (uppermost curve, in red).
\end{center}

\bigskip

A closer look near $\alpha=0$ is provided by the next graph.%
\begin{center}
\includegraphics[
height=3.5994in,
width=5.4106in
]%
{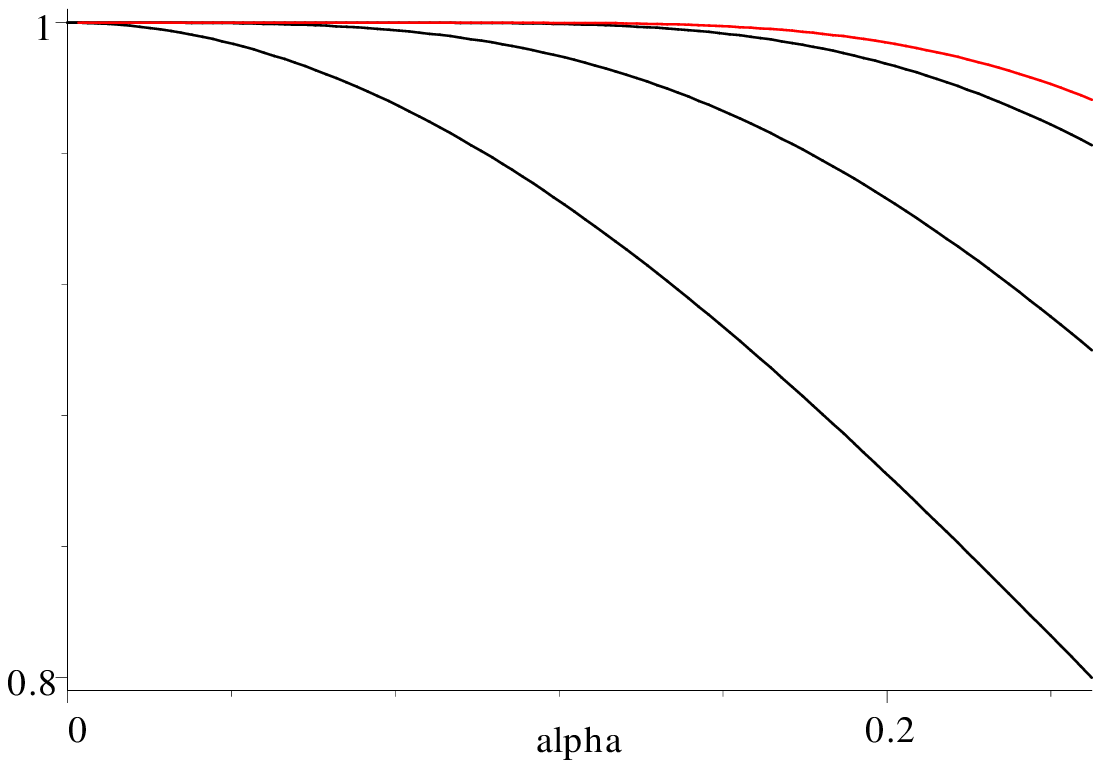}%
\\
$\left.  \mathfrak{B}_{1}^{\left[  j\right]  }\left(  \alpha\right)  \right.
/\alpha=\left.  \mathfrak{B}_{2}^{\left[  j\right]  }\left(  \alpha\right)
\right.  /\alpha^{2}$ for $j=1,2,$ \& $8$ (lower to upper black curves) and
the $j\rightarrow\infty$ limit (uppermost curve, in red).
\end{center}
\newpage

I also plot the corresponding results for $\mathfrak{B}_{3}^{\left[  j\right]
}\left(  \alpha\right)  $ and $\mathfrak{B}_{4}^{\left[  j\right]  }\left(
\alpha\right)  $.
\begin{center}
\includegraphics[
height=3.5994in,
width=5.4106in
]%
{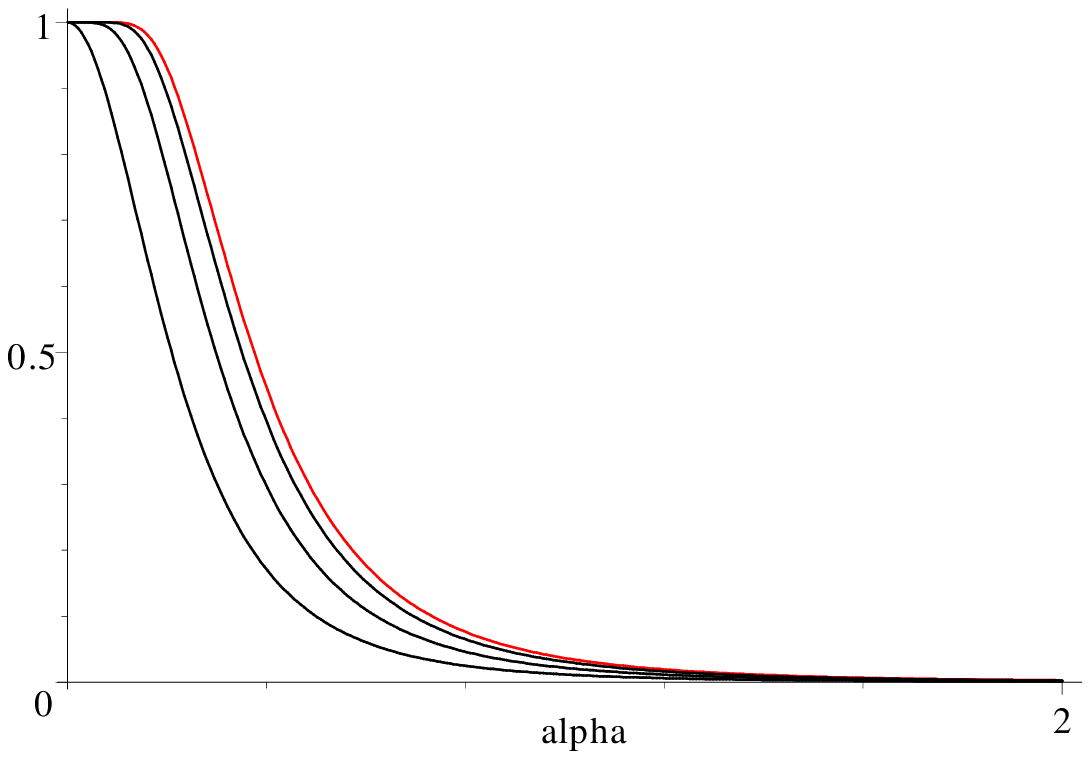}%
\\
$\left.  \mathfrak{B}_{3}^{\left[  j\right]  }\left(  \alpha\right)  \right.
/\alpha^{3}=\left.  \mathfrak{B}_{4}^{\left[  j\right]  }\left(
\alpha\right)  \right.  /\alpha^{4}$ for $j=2,4,$ \& $12$ (lower to upper
black curves) and the $j\rightarrow\infty$ limit (uppermost curve, in red).
\end{center}

\bigskip

Again, a closer look near $\alpha=0$ is given in the next graph.%
\begin{center}
\includegraphics[
height=3.5994in,
width=5.4106in
]%
{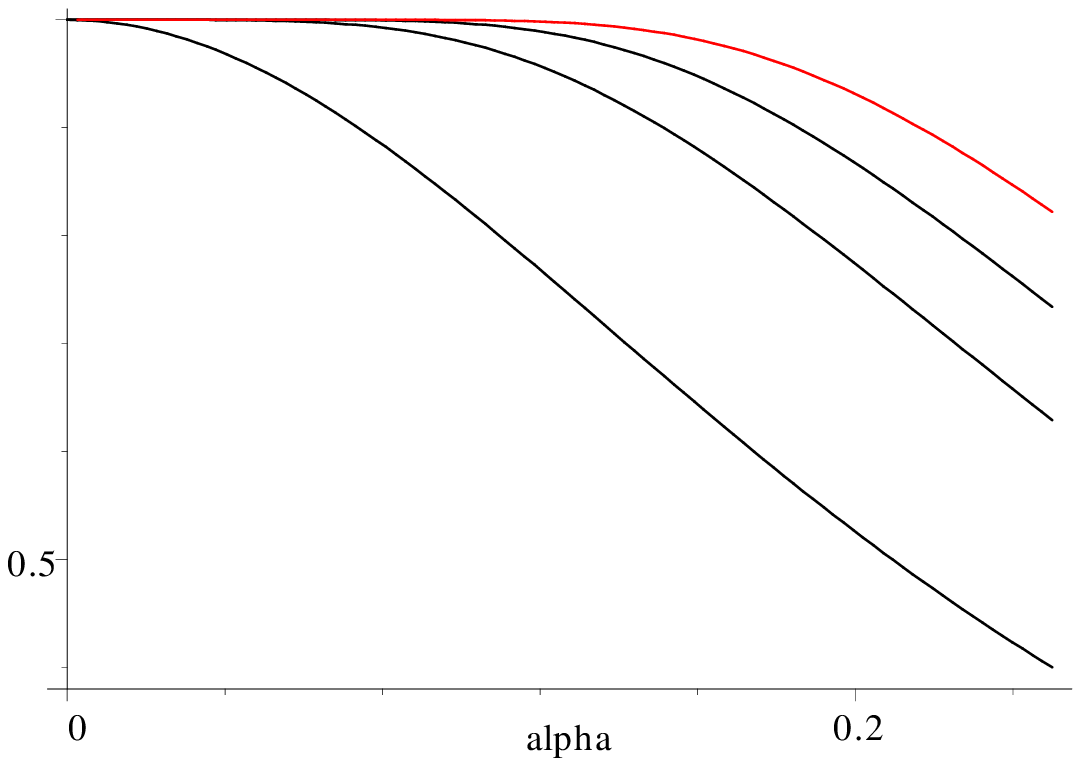}%
\\
$\left.  \mathfrak{B}_{3}^{\left[  j\right]  }\left(  \alpha\right)  \right.
/\alpha^{3}=\left.  \mathfrak{B}_{4}^{\left[  j\right]  }\left(
\alpha\right)  \right.  /\alpha^{4}$ for $j=2,4,$ \& $12$ (lower to upper
black curves) and the $j\rightarrow\infty$ limit (uppermost curve, in red).
\end{center}

\bigskip

Similar behavior is exhibited by the higher coefficients.\newpage

These plots should be compared to the large $j$ behavior of the coefficients
for the exponential, given in (\ref{exp coefficients}). \ As the Figures in
Section 3 attest, the asymptotic $j\rightarrow\infty$ behavior is not closely
approached by the $A_{k}^{\left[  j\right]  }\left(  \theta\right)  $\ for
fixed $k$ until rather large values of $j$ are considered, and for some
$\theta$ not even then. \ This is especially significant for $\theta$ near
$\pm\pi$, where $\lim_{j\rightarrow\infty}A_{k}^{\left[  j\right]  }\left(
\theta\right)  $ has a discontinuity in $\theta$ for every other value of $k$.
\ By contrast, for fixed $k$ as $j\rightarrow\infty$ the $\mathfrak{A}%
_{k}^{\left[  j\right]  }\left(  \alpha\right)  $ converge uniformly and
monotonically to $\mathfrak{A}_{k}^{\left[  \infty\right]  }\left(
\alpha\right)  $\ on any finite interval in $\alpha$, and indeed, the large
$j$ limit of the coefficients is closely approached for moderate values of
$j$. \ These last points are clearly displayed in plots of the relative error
between $\mathfrak{A}_{k}^{\left[  j\right]  }\left(  \alpha\right)  $ and
$\mathfrak{A}_{k}^{\left[  \infty\right]  }\left(  \alpha\right)  $ for a few
values of $j$. \ Further numerical studies suggest that, for any fixed $k$,
the relative error
\begin{equation}
\Delta_{k}^{\left[  j\right]  }\left(  \alpha\right)  \equiv\frac{\left.
\mathfrak{A}_{k}^{\left[  \infty\right]  }\left(  \alpha\right)  \right.
-\left.  \mathfrak{A}_{k}^{\left[  j\right]  }\left(  \alpha\right)  \right.
}{\left.  \mathfrak{A}_{k}^{\left[  j\right]  }\left(  \alpha\right)  \right.
}%
\end{equation}
is always $\geq0$ and goes monotonically to zero as $j\rightarrow\infty$, for
all $\alpha$.

Here are some examples.\
\begin{center}
\includegraphics[
height=3.1506in,
width=4.7223in
]%
{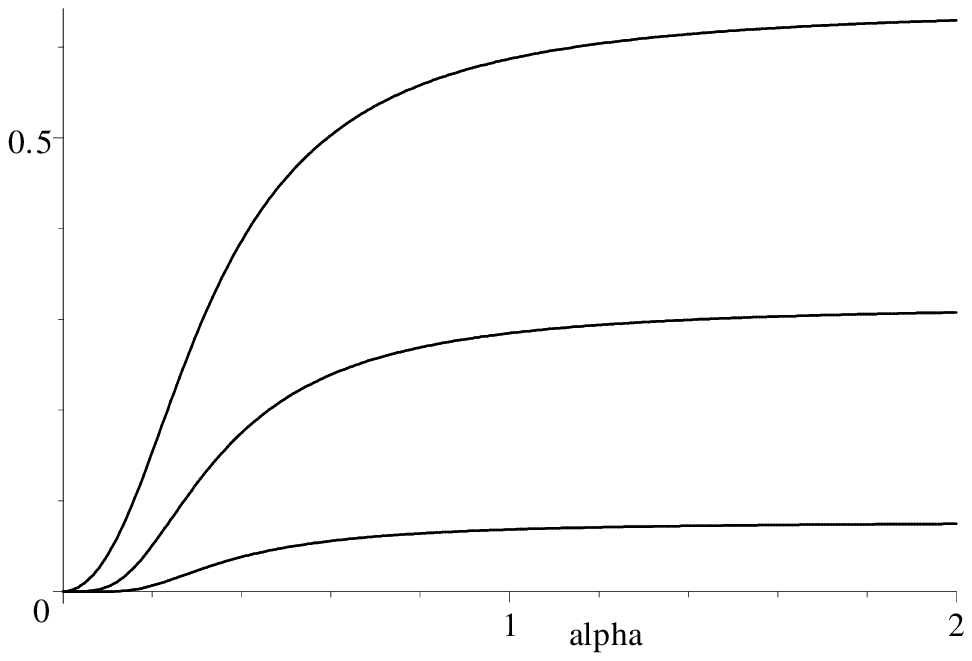}%
\\
The relative error $\Delta_{1}^{\left[  j\right]  }\left(  \alpha\right)
=\Delta_{2}^{\left[  j\right]  }\left(  \alpha\right)  $ plotted for $j=1,2,$
\& $8$, as upper to lower curves, respectively.
\end{center}
\begin{center}
\includegraphics[
height=3.1506in,
width=4.7223in
]%
{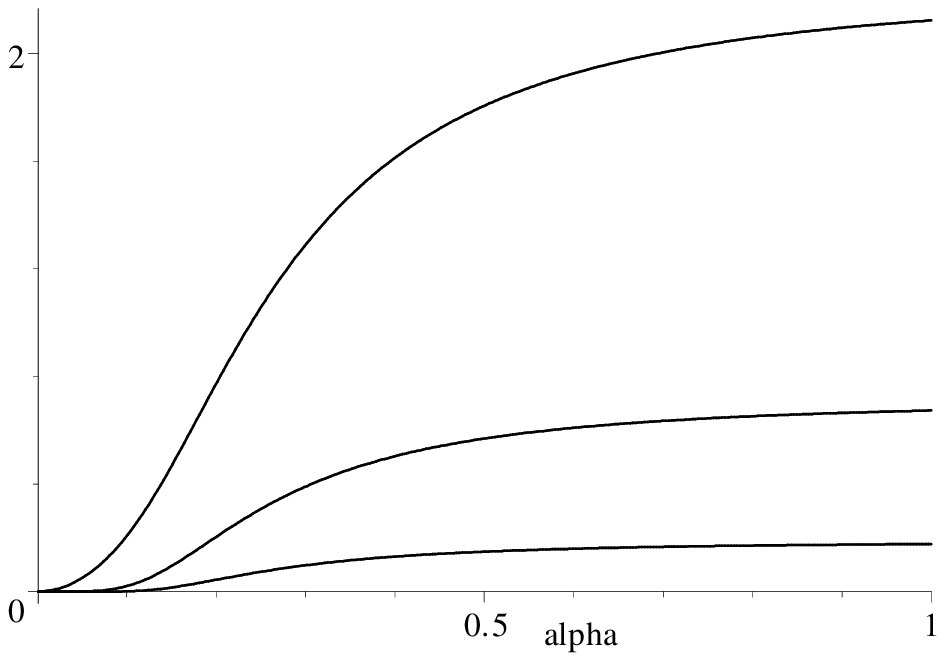}%
\\
The relative error $\Delta_{3}^{\left[  j\right]  }\left(  \alpha\right)
=\Delta_{4}^{\left[  j\right]  }\left(  \alpha\right)  $ plotted for $j=1,2,$
\& $12$, as upper to lower curves, respectively.
\end{center}

Again, similar behavior is exhibited by the higher coefficients. \ 

\subsubsection*{Large $j$ conjectures and proofs for all the coefficients}

Here are the first four requisite asymptotic $j\rightarrow\infty$ results, for
integer $j$, as obtained by explicit calculation:%
\begin{align}
\operatorname*{asymp}\nolimits_{1}\left(  \alpha\right)   &  =1-\frac{1}%
{\frac{2\alpha}{\pi}~\sinh\frac{\pi}{2\alpha}}\ ,\\
\operatorname*{asymp}\nolimits_{2}\left(  \alpha\right)   &  =1-\frac{1}%
{\frac{2\alpha}{\pi}~\sinh\frac{\pi}{2\alpha}}\left(  1+\frac{1}{3!}\left(
\frac{\pi}{2\alpha}\right)  ^{2}\right)  \ ,\\
\operatorname*{asymp}\nolimits_{3}\left(  \alpha\right)   &  =1-\frac{1}%
{\frac{2\alpha}{\pi}~\sinh\frac{\pi}{2\alpha}}\left(  1+\frac{1}{3!}\left(
\frac{\pi}{2\alpha}\right)  ^{2}+\frac{1}{5!}\left(  \frac{\pi}{2\alpha
}\right)  ^{4}\right)  \ ,\\
\operatorname*{asymp}\nolimits_{4}\left(  \alpha\right)   &  =1-\frac{1}%
{\frac{2\alpha}{\pi}~\sinh\frac{\pi}{2\alpha}}\left(  1+\frac{1}{3!}\left(
\frac{\pi}{2\alpha}\right)  ^{2}+\frac{1}{5!}\left(  \frac{\pi}{2\alpha
}\right)  ^{4}+\frac{1}{7!}\left(  \frac{\pi}{2\alpha}\right)  ^{6}\right)
\ .
\end{align}
(Semi-integer $j$ are different. \ See below.) \ Note that the numerators are
just truncations of the series for the denominator, $\frac{2\alpha}{\pi}%
~\sinh\frac{\pi}{2\alpha}=\sum_{n=0}^{\infty}\frac{1}{\left(  2n+1\right)
!}\left(  \frac{\pi}{2\alpha}\right)  ^{2n}$. \ Hence the obvious conjecture
for integer $j$ is%
\begin{equation}
\operatorname*{asymp}\nolimits_{k}\left(  \alpha\right)  =1-\frac{1}%
{\frac{2\alpha}{\pi}~\sinh\frac{\pi}{2\alpha}}\left(  \sum_{n=0}^{k-1}\frac
{1}{\left(  2n+1\right)  !}\left(  \frac{\pi}{2\alpha}\right)  ^{2n}\right)
=\frac{\left(  \frac{\pi}{2\alpha}\right)  ^{1+2k}}{\left(  1+2k\right)
!\sinh\frac{\pi}{2\alpha}}~\operatorname{hypergeom}\left(  \left[  1\right]
,\left[  1+k,\frac{3}{2}+k\right]  ,\frac{\pi^{2}}{16\alpha^{2}}\right)  \ .
\label{AsympK}%
\end{equation}
Given this result, the coefficients of the spin matrix powers in the reduction
(\ref{GeometricSeriesPolynomial}) are, as $j\rightarrow\infty$ \textit{for
integer }$j$,%
\begin{equation}
\lim_{j\rightarrow\infty}\frac{1}{\alpha^{2k-1}}\left.  \mathfrak{B}%
_{2k-1}^{\left[  j\right]  }\left(  \alpha\right)  \right.  =\lim
_{j\rightarrow\infty}\frac{1}{\alpha^{2k}}\left.  \mathfrak{B}_{2k}^{\left[
j\right]  }\left(  \alpha\right)  \right.  =1-\frac{1}{\frac{2\alpha}{\pi
}~\sinh\frac{\pi}{2\alpha}}\left(  \sum_{n=0}^{k-1}\frac{1}{\left(
2n+1\right)  !}\left(  \frac{\pi}{2\alpha}\right)  ^{2n}\right)  \ ,\text{
\ \ for }k\geq1\ \text{.} \label{AsympBosonic}%
\end{equation}
As an immediate consequence, $\lim_{k\rightarrow\infty}\operatorname*{asymp}%
\nolimits_{k}\left(  \alpha\right)  =0$ for all $\alpha\neq0$, with the limit
approached monotonically from above, while at $\alpha=0,$
$\operatorname*{asymp}\nolimits_{k}\left(  0\right)  =1$ for all $k$.

Also, to put it briefly, explicit calculation of the lowest few coefficients
reveals the corresponding conjecture for semi-integer $j$: \ The asymptotic
expressions may be obtained from the bosonic spin case by substituting
$\frac{2\alpha}{\pi}\sinh\frac{\pi}{2\alpha}\Longrightarrow\cosh\frac{\pi
}{2\alpha}$. \ Thus for the fermionic case, as $j\rightarrow\infty$
\textit{for semi-integer }$j$,
\begin{equation}
\lim_{j\rightarrow\infty}\frac{1}{\alpha^{2k}}\left.  \mathfrak{B}%
_{2k}^{\left[  j\right]  }\left(  \alpha\right)  \right.  =\lim_{j\rightarrow
\infty}\frac{1}{\alpha^{2k+1}}\left.  \mathfrak{B}_{2k+1}^{\left[  j\right]
}\left(  \alpha\right)  \right.  =1-\frac{1}{\cosh\frac{\pi}{2\alpha}}\left(
\sum_{n=0}^{k}\frac{1}{\left(  2n\right)  !}\left(  \frac{\pi}{2\alpha
}\right)  ^{2n}\right)  \ ,\text{ \ \ for }k\geq0\ \text{.}
\label{AsympFermionic}%
\end{equation}
While plots of the various functions of $\alpha$ in (\ref{AsympBosonic}) and
(\ref{AsympFermionic}) are qualitatively similar, nevertheless, the bosonic
and fermionic Cayley transforms are still easily distinguished, even as
$j\rightarrow\infty$.

Proofs of these conjectures are straightforward using (\ref{GSPCoefficients}),
(\ref{TruncsAsCFNs}), (\ref{AsymptoticClosedFormDeterminants}), and
Proposition 2.7 in \cite{CFN}. \ For example, for integer $j$, use the result
(\ref{TruncsAsCFNs}) to rewrite (\ref{GSPCoefficients}) for the even-indexed
coefficients as
\begin{equation}
1-\frac{1}{\alpha^{2k}}\left.  \mathfrak{B}_{2k}^{\left[  j\right]  }\left(
\alpha\right)  \right.  =\frac{1}{\det\left(  1-2i\alpha~\boldsymbol{\hat
{n}\cdot J}\right)  }~\sum_{l=1}^{k}\left(  2\alpha\right)  ^{2\left(
j+1-l\right)  }~\left\vert t\left(  2j+2,2l\right)  \right\vert \ .
\end{equation}
The limit $j\rightarrow\infty$ now follows from the integer $j$ result in
(\ref{AsymptoticClosedFormDeterminants}), written as
\begin{equation}
\lim_{j\rightarrow\infty}\left(  \frac{\left(  2\alpha\right)  ^{2j}\left(
\Gamma\left(  1+j\right)  \right)  ^{2}}{\det\left(  1-2i\alpha
~\boldsymbol{\hat{n}\cdot J}\right)  }\right)  =\frac{\pi}{2\alpha\sinh\left(
\frac{\pi}{2\alpha}\right)  }\ ,
\end{equation}
and from the asymptotic form of $\left\vert t\left(  2j+2,2m\right)
\right\vert $ as given in \cite{CFN}, Proposition 2.7 (xxvi), written as\
\begin{equation}
\lim_{j\rightarrow\infty}\frac{\left(  2\alpha\right)  ^{2\left(  1-l\right)
}}{\left(  \Gamma\left(  1+j\right)  \right)  ^{2}}~\left\vert t\left(
2j+2,2l\right)  \right\vert =\frac{1}{\left(  2l-1\right)  !}\left(  \frac
{\pi}{2\alpha}\right)  ^{2\left(  l-1\right)  }\ .
\end{equation}
The net result is an expression whose numerator is a truncation of the
infinite series for the denominator, as conjectured above.%
\begin{equation}
\lim_{j\rightarrow\infty}\left(  1-\frac{1}{\alpha^{2k}}\left.  \mathfrak{B}%
_{2k}^{\left[  j\right]  }\left(  \alpha\right)  \right.  \right)  =\frac
{1}{\frac{2\alpha}{\pi}~\sinh\frac{\pi}{2\alpha}}\left(  \sum_{n=0}^{k-1}%
\frac{1}{\left(  2n+1\right)  !}\left(  \frac{\pi}{2\alpha}\right)
^{2n}\right)  \ .
\end{equation}
The odd-indexed coefficients for the bosonic spin case are given by this same
expression according to the elementary pairing (\ref{BosonicPairing}). \ Thus
for integer $j$ the conjectured $j\rightarrow\infty$\ form of the coefficients
is established.

A corresponding proof for semi-integer $j$ follows from
(\ref{AsymptoticClosedFormDeterminants}) and Proposition 2.7 (xxvii) in
\cite{CFN}. \ \newpage

The result for semi-integer $j$\ is as conjectured in (\ref{AsympFermionic}).
\ Alternatively, it may be expressed as a hypergeometric function similar to
that in (\ref{AsympK}).%
\begin{equation}
1-\frac{1}{\cosh\frac{\pi}{2\alpha}}\left(  \sum_{n=0}^{k}\frac{1}{\left(
2n\right)  !}\left(  \frac{\pi}{2\alpha}\right)  ^{2n}\right)  =\frac{\left(
\frac{\pi}{2\alpha}\right)  ^{2+2k}}{\left(  2+2k\right)  !\cosh\frac{\pi
}{2\alpha}}~\operatorname{hypergeom}\left(  \left[  1\right]  ,\left[
\frac{3}{2}+k,2+k\right]  ,\frac{\pi^{2}}{16\alpha^{2}}\right)  \ .
\end{equation}

\section{Relating exponentials and rational forms by variable changes}

To finish the comparison of rotations as exponentials and Cayley transforms,
for arbitrary $j$, it is instructive to return to an issue briefly mentioned
in the Introduction. \ Namely, why does a map from one form to the other
\textbf{not} succeed through the use of a numerical-valued change of variable
$\alpha\left(  \theta\right)  $ or its inverse $\theta\left(  \alpha\right)
$? \ After all, a comparison of the exponential and Cayley results for
$j=1/2$, as well as for $j=1$, as given in (\ref{ExpSpinHalf}),
(\ref{ExpSpinOne}), (\ref{CayleySpinHalf}), and (\ref{CayleySpinOne}), would
give the \emph{misleading} impression that this can succeed.

To see the difficulty, take a basis where $\boldsymbol{\hat{n}\cdot J}$\ is
diagonal, with eigenvalues $M\in\left\{  -j,-j+1,\cdots,j-1,j\right\}  $, and
equate the exponential and Cayley forms acting on a specific eigenstate
$\left\vert M\right\rangle $. \ The result of course is%
\begin{equation}
\exp\left(  i~\theta~M\right)  =\frac{1+2i\alpha~M}{1-2i\alpha~M}\ .
\label{ExpEqualCayley}%
\end{equation}
This is easily solved either for $\alpha\left(  \theta\right)  $ or for the
inverse map $\theta\left(  \alpha\right)  $:%
\begin{equation}
\alpha\left(  \theta\right)  =\frac{1}{2M}\tan\left(  \frac{M\theta}%
{2}\right)  \ ,\ \ \ \theta\left(  \alpha\right)  =\frac{2}{M}~\arctan\left(
2M\alpha\right)  \ . \label{MDependentMaps}%
\end{equation}
So the relation is $M$-dependent, although in this particular instance it
\emph{is} \emph{in}dependent of the sign of $M$. \ In general, nontrivial
$M$-dependence would be a feature that would arise upon comparing \emph{any}
two functions of $\boldsymbol{\hat{n}\cdot J}$ when acting on $\left\vert
M\right\rangle $.

Now it just so happens that a unique map works for \emph{all} the
$M$-eigenstates for $j=1/2$, and another unique map works for all the
$M$-eigenstates states for $j=1$. \ (Note that for $M=0$ \ the relation
(\ref{ExpEqualCayley}) always reduces to $1=1$ \emph{no matter what} the
relation between $\alpha$ and $\theta$.) \ But this is fortuitous and peculiar
to those two spins due to $\left\vert M\right\vert $\ having only one
nontrivial value in those special cases. \ For all $j\geq3/2$ the map must
change with $\left\vert M\right\vert $. \ There is \emph{no consistent
}$\alpha\rightleftarrows\theta$\emph{ relation} that works on all
$\boldsymbol{\hat{n}\cdot J}$ eigenstates for $j>1$. \ In a sense made precise
by (\ref{MDependentMaps}), there is a \textquotedblleft parameter
shear\textquotedblright\ in the relation between $\alpha$ and $\theta$ as the
$\boldsymbol{\hat{n}\cdot J}$ spectrum is traversed.

Similarly, if one were to equate the coefficients of any given power $\left(
\boldsymbol{\hat{n}\cdot J}\right)  ^{k}$ in the spin matrix polynomial
reductions of the exponential and the Cayley transform, the result would be an
$\alpha\rightleftarrows\theta$ relation that depends on $k$, and this result
would be \emph{in}consistent (except in the two special cases $j=1/2$ and
$j=1$) with that obtained from equating the coefficients of some other power
of $\boldsymbol{\hat{n}\cdot J}$.

\section{Relating exponentials and rational forms by Laplace transforms}

Some standard results from
\href{https://en.wikipedia.org/wiki/Resolvent_formalism}{resolvent theory}
provide a direct relation between (\ref{the result}) and
(\ref{GeometricSeriesPolynomial}). \ Consider the
\href{https://en.wikipedia.org/wiki/Laplace_transform}{Laplace transform} and
its inverse,%
\begin{equation}
F\left(  s\right)  =\mathcal{L}\left[  f\right]  =\int_{0}^{\infty}%
e^{-st}f\left(  t\right)  dt\ ,\ \ \ f\left(  t\right)  =\mathcal{L}%
^{-1}\left[  F\right]  =\frac{1}{2\pi i}\lim_{T\rightarrow\infty}\int%
_{\gamma-iT}^{\gamma+iT}e^{st}F\left(  s\right)  ds\ .
\end{equation}
For any hermitian $N\times N$ matrix, a Laplace transform yields the resolvent
from the exponential, and vice versa.%
\begin{equation}
\frac{1}{s-iM}=\mathcal{L}\left[  e^{itM}\right]  \ ,\ \ \ e^{itM}%
=\mathcal{L}^{-1}\left[  \frac{1}{s-iM}\right]  \ . \label{HY}%
\end{equation}
This furnishes another route to derive the CFZ formula starting from the known
matrix polynomial expression for the resolvent. \ Given the ease by which the
polynomial coefficients for the resolvent are obtained, this provides perhaps
the simplest proof of the CFZ result.

For linearly independent powers $M^{k}$, $0\leq k<N$, as is the case for spin
matrices, the matrix polynomial expansion coefficients for the exponential and
the resolvent are distinctly related order-by-order by the Laplace transform.%
\begin{equation}
\exp\left(  itM\right)  =\sum_{n=0}^{N-1}A_{n}\left(  t\right)  \left(
iM\right)  ^{n}\ ,\ \ \ \frac{1}{s-iM}=\sum_{n=0}^{N-1}B_{n}\left(  s\right)
\left(  iM\right)  ^{n}\ ,\ \ \ B_{n}\left(  s\right)  =\mathcal{L}\left[
A_{n}\right]  \ ,\ \ \ A_{n}\left(  t\right)  =\mathcal{L}^{-1}\left[
B_{n}\right]  \ . \label{HYCoeffs}%
\end{equation}
If the first $k$ powers of $M$ are not independent, for $k<N-1$, correct
relations between the exponential and the resolvent are still obtained by
taking $\mathcal{L}$\ or $\mathcal{L}^{-1}$, as in (\ref{HY}) and
(\ref{HYCoeffs}), although the final results may admit further simplification.
\ Fortunately, it is not necessary to worry about that situation in the following.

Rescaling variables produces a form of the Laplace transform that is more
useful for the direct comparison of (\ref{exp coefficients})\ and
(\ref{GSPCoefficients}). \ Thus%
\begin{align}
\frac{1}{1-2i~\alpha~\boldsymbol{\hat{n}\cdot J}}  &  =\int_{0}^{\infty}%
e^{-t}\exp\left(  2i~\alpha t~\boldsymbol{\hat{n}\cdot J}\right)  ~dt\ ,\\
\left.  \mathfrak{B}_{k}^{\left[  j\right]  }\left(  \alpha\right)  \right.
&  =\frac{1}{k!}\int_{0}^{\infty}e^{-t}\left.  A_{k}^{\left[  j\right]
}\left(  2\alpha t\right)  \right.  dt\ . \label{BasLaplaceTransformOfA}%
\end{align}
It suffices here to consider only the direct transform. \ The inverse relation
is automatic, given the large $\alpha$ behavior of the $\left.  \mathfrak{B}%
_{k}^{\left[  j\right]  }\left(  \alpha\right)  \right.  $. \ From
(\ref{the coefficients as cfns}), along with (\ref{oddbydiff}),\ the required
calculation is
\begin{equation}
I_{m}^{\left[  j\right]  }\left(  \alpha\right)  =\frac{1}{m!}\int_{0}%
^{\infty}e^{-t}\sin^{m}\left(  \alpha t\right)  dt\text{ \ \ for
even\ }2j-k\ .
\end{equation}
These are elementary integrals, of course, resulting in%
\begin{equation}
I_{m}^{\left[  j\right]  }\left(  \alpha\right)  =\alpha^{m}%
{\displaystyle\prod\limits_{l=1}^{m/2}}
\frac{1}{1+4l^{2}\alpha^{2}}\text{ \ \ for even }m\ ,\ \ \ I_{m}^{\left[
j\right]  }\left(  \alpha\right)  =\alpha^{m}%
{\displaystyle\prod\limits_{l=1}^{\left(  m+1\right)  /2}}
\frac{1}{1+\left(  2l-1\right)  ^{2}\alpha^{2}}\text{ \ \ for odd }m\ .
\end{equation}
After combining these results with (\ref{the coefficients as cfns}) and
(\ref{Determinant}), and some elementary manipulations of the central
factorial numbers, the transform (\ref{BasLaplaceTransformOfA})\ yields the
expected result:%
\begin{equation}
\mathfrak{B}_{k}^{\left[  j\right]  }\left(  \alpha\right)  =\frac{\alpha^{k}%
}{\det\left(  1-2i\alpha~\boldsymbol{\hat{n}\cdot J}\right)  }~\sum
_{m=0}^{\left\lfloor j-k/2\right\rfloor }4^{m}\alpha^{2m}~\left\vert t\left(
2j+2,2j+2-2m\right)  \right\vert \ .
\end{equation}

For example, for spin $j=1/2$, with the usual $2\times2$ matrices,%
\begin{gather}
\exp\left(  i~\theta~\boldsymbol{\hat{n}\cdot J}\right)  =\cos\left(
\theta/2\right)  ~\boldsymbol{I}+2i\sin\left(  \theta/2\right)  ~\left(
\boldsymbol{\hat{n}\cdot J}\right)  \ ,\\
\nonumber\\
\int_{0}^{\infty}e^{-t}\cos\left(  \alpha t\right)  dt=\frac{1}{1+\alpha^{2}%
}\ ,\ \ \ \int_{0}^{\infty}e^{-t}\sin\left(  \alpha t\right)  dt=\frac{\alpha
}{1+\alpha^{2}}\ ,\\
\nonumber\\
\det\left(  1-2i\alpha~\boldsymbol{\hat{n}\cdot J}\right)  =1+\alpha^{2}\ ,\\
\nonumber\\
\frac{1}{1-2i\alpha~\boldsymbol{\hat{n}\cdot J}}=\frac{1}{1+\alpha^{2}}\left(
\boldsymbol{I}+2i\alpha~\boldsymbol{\hat{n}\cdot J}\right)  \ ,
\end{gather}
and for spin $j=1$, with $3\times3$ matrices,%
\begin{gather}
\exp\left(  i~\theta~\boldsymbol{\hat{n}\cdot J}\right)  =\boldsymbol{I}%
+i\sin\left(  \theta\right)  \boldsymbol{\left(  \boldsymbol{\hat{n}\cdot
J}\right)  +}\left(  \cos\theta-1\right)  \left(  \boldsymbol{\hat{n}\cdot
J}\right)  ^{2}\ ,\\
\nonumber\\
\int_{0}^{\infty}e^{-t}dt=1\ ,\ \ \ \int_{0}^{\infty}e^{-t}\cos\left(  2\alpha
t\right)  dt=\frac{1}{1+4\alpha^{2}}\ ,\ \ \ \int_{0}^{\infty}e^{-t}%
\sin\left(  2\alpha t\right)  dt=\frac{2\alpha}{1+4\alpha^{2}}\\
\nonumber\\
\det\left(  1-2i\alpha~\boldsymbol{\hat{n}\cdot J}\right)  =1+4\alpha^{2}\ ,\\
\nonumber\\
\frac{1}{1-2i\alpha~\boldsymbol{\hat{n}\cdot J}}=\frac{1}{1+4\alpha^{2}%
}\left(  \left(  1+4\alpha^{2}\right)  I+2i\alpha~\boldsymbol{\hat{n}\cdot
J-}4\alpha^{2}\left(  \boldsymbol{\hat{n}\cdot J}\right)  ^{2}\right)  \ .
\end{gather}
\newpage

\section{Concluding remarks}

Polynomial reductions of spin matrix exponentials have long been recognized as
important and useful in many different contexts \cite{Illamed}-\cite{T},
\cite{W}-\cite{Nelson}. \ Perhaps the results presented in \cite{CFZ,CvK,TSvK}%
\ and discussed further in this paper can help to facilitate these and other applications.

Moreover, Cayley transforms have numerous applications, many of a practical
nature. \ For example, consider \cite{TJS} and \cite{I}. \ In the first of
these two papers, previous results on attitude representations were
generalized using Cayley transforms, for application to guidance and control
problems, while in the second paper, the theory of time stepping methods were
developed in a systematic manner based on the Cayley transform, for
application to discretised differential equations that describe evolution in
Lie groups. Both of these papers stress that Cayley transform methods are
easier to implement than the evaluation of matrix exponentials. \ 

This relative ease of implementation is borne out in the present paper by
direct comparison of (\ref{exp coefficients})\ and (\ref{GSPCoefficients}),
and more indirectly by the relative simplicity of the derivation for the
Cayley transform coefficients, as given here, when compared to the proofs for
the coefficients in the exponential case, as given in \cite{CvK}. \ 

\paragraph*{Acknowledgements}

I thank T S Van Kortryk and C K Zachos for discussions. \ I have also
benefited from stimulating views of sea waves approaching the Malec\'{o}n, La
Habana, and I thank the organizers of STARS 2015 for their efforts to make
this possible. \ This work was supported in part by NSF Award PHY-1214521, and
in part by a University of Miami Cooper Fellowship.

\subsection*{Appendix A: \ Central factorial numbers}

For historical reasons, central factorial numbers are defined as the
coefficients in simple polynomials \cite{Riordan,CFN}. \ They can be either
positive or negative, but only their absolute values are needed for the
coefficients of the spin matrix expansions in the main text. \ Moreover,
$t\left(  even,even\right)  $ are integers, but $t\left(  odd,odd\right)  $
are not integers, and $t\left(  odd,even\right)  =0=t\left(  even,odd\right)
$. \ So the even and odd cases are best handled separately. \ 

By definition and as elementary consequences thereof (cf.
\emph{http://oeis.org/A182867}),%
\begin{align}
\prod\limits_{l=0}^{m-1}\left(  x^{2}-l^{2}\right)   &  =\sum_{k=1}%
^{m}t\left(  2m,2k\right)  x^{2k}\ ,\tag{A1}\label{A1}\\
t\left(  2m,2k\right)   &  =\left(  -1\right)  ^{m-k}\left\vert t\left(
2m,2k\right)  \right\vert =\left(  -1\right)  ^{m-k}\frac{1}{k!}\left.
\frac{d^{k}}{dz^{k}}\prod\limits_{l=0}^{m-1}\left(  z+l^{2}\right)
\right\vert _{z=0}\ , \tag{A2}\label{A2}%
\end{align}
as well as (cf. \emph{http://oeis.org/A008956})%
\begin{align}
x\prod\limits_{l=0}^{m-1}\left(  x^{2}-\left(  l+\frac{1}{2}\right)
^{2}\right)   &  =\sum_{k=0}^{m}t\left(  2m+1,2k+1\right)  x^{2k+1}%
\ ,\tag{A3}\label{A3}\\
t\left(  2m+1,2k+1\right)   &  =\left(  -1\right)  ^{m-k}\left\vert t\left(
2m+1,2k+1\right)  \right\vert =\left(  -1\right)  ^{m-k}\frac{1}{k!}\left.
\frac{d^{k}}{dz^{k}}\prod\limits_{l=0}^{m-1}\left(  z+\left(  l+\frac{1}%
{2}\right)  ^{2}\right)  \right\vert _{z=0}\ . \tag{A4}\label{A4}%
\end{align}

\subsection*{Appendix B: \ Proof of the fundamental identity}

Here I provide a demonstration that%
\begin{equation}
\left(  2\boldsymbol{\hat{n}\cdot J}\right)  ^{2j+1}=-\sum_{k=0}%
^{2j}2^{1+2j-k}\times t\left(  2+2j,1+k\right)  \times\left(
2\boldsymbol{\hat{n}\cdot J}\right)  ^{k}\ , \tag{B1}\label{LemmaResult}%
\end{equation}
where $t\left(  m,n\right)  $ are the central factorial numbers, defined in
Appendix A. \ In a basis where $2\boldsymbol{\hat{n}\cdot J}$ is diagonal,
(\ref{LemmaResult}) reduces to a matrix equation,
\begin{equation}
\left[
\begin{array}
[c]{c}%
\left(  2j\right)  ^{2j+1}\smallskip\\
\left(  2j-2\right)  ^{2j+1}\\
\vdots\smallskip\\
\left(  -2j+2\right)  ^{2j+1}\smallskip\\
\left(  -2j\right)  ^{2j+1}%
\end{array}
\right]  =-2^{1+2j}\times V\left[  j\right]  \left[
\begin{array}
[c]{c}%
t\left(  2+2j,1\right)  \smallskip\\
\frac{1}{2}~t\left(  2+2j,2\right) \\
\vdots\smallskip\\
\frac{1}{2^{2j-1}}~t\left(  2+2j,2j\right)  \smallskip\\
\frac{1}{2^{2j}}~t\left(  2+2j,1+2j\right)
\end{array}
\right]  \ , \tag{B2}\label{LemmaRHS}%
\end{equation}
where the Vandermonde matrix for spin $j$ is defined as in (\ref{Vandermonde}%
). \ So, consider the $k$th row on the RHS of (\ref{LemmaRHS}): \
\begin{equation}
-\frac{2^{1+2j}}{\left(  j+1-k\right)  }\times\left(
\begin{array}
[c]{c}%
t\left(  2+2j,1\right)  \left(  j+1-k\right)  +t\left(  2+2j,2\right)  \left(
j+1-k\right)  ^{2}+t\left(  2+2j,3\right)  \left(  j+1-k\right)  ^{3}+\cdots\\
+t\left(  2+2j,2j\right)  \left(  j+1-k\right)  ^{2j-1}+t\left(
2+2j,1+2j\right)  \left(  j+1-k\right)  ^{2j}%
\end{array}
\right)  \ . \tag{B3}%
\end{equation}
If $j$ is semi-integer, say $j=n+1/2$ for integer $n$, then $t\left(
2+2j,even\right)  =0$, and this $k$th row becomes%
\begin{align}
&  -\frac{2^{1+2j}}{\left(  j+1-k\right)  }\times\left(
\begin{array}
[c]{c}%
t\left(  2+2j,1\right)  \left(  j+1-k\right)  +t\left(  2+2j,3\right)  \left(
j+1-k\right)  ^{3}+\cdots\\
+t\left(  2+2j,2j-2\right)  \left(  j+1-k\right)  ^{2j-2}+t\left(
2+2j,2j\right)  \left(  j+1-k\right)  ^{2j}\\
+t\left(  2+2j,2j+2\right)  \left(  j+1-k\right)  ^{2j+2}-t\left(
2+2j,2j+2\right)  \left(  j+1-k\right)  ^{2j+2}%
\end{array}
\right) \nonumber\\
&  =-\frac{2^{1+2j}}{\left(  j+1-k\right)  }\times\left(  \left(
j+1-k\right)  \prod\limits_{l=0}^{n}\left(  \left(  j+1-k\right)  ^{2}-\left(
l+\frac{1}{2}\right)  ^{2}\right)  -t\left(  2+2j,2j+2\right)  \left(
j+1-k\right)  ^{2j+2}\right)  \text{\ ,} \tag{B4}%
\end{align}
where the $t\left(  2+2j,2j+2\right)  $ term was added and subtracted to
obtain the complete sum on the RHS of (\ref{A3}), for $m=j+1/2=n+1$ and
$x=j+1-k$. \ The sum was then replaced with the product on the LHS of
(\ref{A3}). \ But the product evaluates to zero because one of the terms in
the product always vanishes for $k\geq1$, and therefore the $k$th row on the
RHS of (\ref{LemmaRHS}) is%
\begin{equation}
t\left(  2+2j,2+2j\right)  \times\left(  2j+2-2k\right)  ^{2j+1}=\left(
2j+2-2k\right)  ^{2j+1}\ , \tag{B5}%
\end{equation}
since\ $t\left(  2+2j,2j+2\right)  =1$. \ Thus the $k$th row on the LHS of
(\ref{LemmaRHS}) is obtained, and the identity is established for semi-integer
$j$. \ If $j$ is integer, a similar proof goes through upon using (\ref{A1}).
\ (See \cite{CvK}, Appendix B.)

\subsection*{Appendix C: \ Biorthogonal matrix examples}

Here are more details about the biorthogonal systems of spin matrices
described in the main text, for $j=1/2,\ 1,\ 3/2,$ and $2$.

Spin $j=1/2$ is deceptively simple. \ The independent powers of the spin
matrix are%
\begin{equation}
S^{0}=\left[
\begin{array}
[c]{cc}%
1 & 0\\
0 & 1
\end{array}
\right]  \ ,\ \ \ S^{1}=\left[
\begin{array}
[c]{cc}%
1 & 0\\
0 & -1
\end{array}
\right]  \ , \tag{C1}%
\end{equation}
and the corresponding trace-orthonormal dual matrices are the same up to a
normalization factor. (NB This is \emph{not} true for any other $j$.)%
\begin{equation}
T_{0}=\frac{1}{2}\left[
\begin{array}
[c]{cc}%
1 & 0\\
0 & 1
\end{array}
\right]  \ ,\ \ \ T_{1}=\frac{1}{2}\left[
\begin{array}
[c]{cc}%
1 & 0\\
0 & -1
\end{array}
\right]  \ ,\ \ \ \text{i.e. \ \ }V^{-1}=\frac{1}{2}\left[
\begin{array}
[c]{cc}%
1 & 1\\
1 & -1
\end{array}
\right]  \ . \tag{C2}%
\end{equation}

Spin $j=1$ is a more interesting example. \ The independent powers of the spin
matrix are given by%
\begin{equation}
S^{0}=\left[
\begin{array}
[c]{ccc}%
1 & 0 & 0\\
0 & 1 & 0\\
0 & 0 & 1
\end{array}
\right]  \ ,\ \ \ S^{1}=\left[
\begin{array}
[c]{ccc}%
2 & 0 & 0\\
0 & 0 & 0\\
0 & 0 & -2
\end{array}
\right]  \ ,\ \ \ S^{2}=\left[
\begin{array}
[c]{ccc}%
4 & 0 & 0\\
0 & 0 & 0\\
0 & 0 & 4
\end{array}
\right]  \ , \tag{C3}%
\end{equation}
and the corresponding trace-orthonormal dual matrices, as well as $V^{-1}$,
are given by
\begin{equation}
T_{0}=\left[
\begin{array}
[c]{ccc}%
0 & 0 & 0\\
0 & 1 & 0\\
0 & 0 & 0
\end{array}
\right]  \ ,\ \ \ T_{1}=\frac{1}{4}\left[
\begin{array}
[c]{ccc}%
1 & 0 & 0\\
0 & 0 & 0\\
0 & 0 & -1
\end{array}
\right]  \ ,\ \ \ T_{3}=\frac{1}{8}\left[
\begin{array}
[c]{ccc}%
1 & 0 & 0\\
0 & -2 & 0\\
0 & 0 & 1
\end{array}
\right]  \ ,\ \ \ V^{-1}=\frac{1}{8}\left[
\begin{array}
[c]{ccc}%
0 & 8 & 0\\
2 & 0 & -2\\
1 & -2 & 1
\end{array}
\right]  \ . \tag{C4}%
\end{equation}

Spin $j=3/2$ is also interesting. \ The independent powers of the spin matrix
are given by%
\begin{align}
S^{0}  &  =\left[
\begin{array}
[c]{cccc}%
1 & 0 & 0 & 0\\
0 & 1 & 0 & 0\\
0 & 0 & 1 & 0\\
0 & 0 & 0 & 1
\end{array}
\right]  \ ,\ \ \ S^{1}=\left[
\begin{array}
[c]{cccc}%
3 & 0 & 0 & 0\\
0 & 1 & 0 & 0\\
0 & 0 & -1 & 0\\
0 & 0 & 0 & -3
\end{array}
\right]  \ ,\ \ \ S^{2}=\left[
\begin{array}
[c]{cccc}%
9 & 0 & 0 & 0\\
0 & 1 & 0 & 0\\
0 & 0 & 1 & 0\\
0 & 0 & 0 & 9
\end{array}
\right]  \ ,\ \ \ S^{3}=\left[
\begin{array}
[c]{cccc}%
27 & 0 & 0 & 0\\
0 & 1 & 0 & 0\\
0 & 0 & -1 & 0\\
0 & 0 & 0 & -27
\end{array}
\right]  \ ,\nonumber\\
&  \tag{C5}%
\end{align}
and the corresponding trace-orthonormal dual matrices, as well as $V^{-1}$,
are given by%
\begin{gather}
T_{0}=\frac{1}{16}\left[
\begin{array}
[c]{cccc}%
-1 & 0 & 0 & 0\\
0 & 9 & 0 & 0\\
0 & 0 & 9 & 0\\
0 & 0 & 0 & -1
\end{array}
\right]  \ ,\ \ \ T_{1}=\frac{1}{48}\left[
\begin{array}
[c]{cccc}%
-1 & 0 & 0 & 0\\
0 & 27 & 0 & 0\\
0 & 0 & -27 & 0\\
0 & 0 & 0 & 1
\end{array}
\right]  \ ,\ \ \ T_{2}=\frac{1}{16}\left[
\begin{array}
[c]{cccc}%
1 & 0 & 0 & 0\\
0 & -1 & 0 & 0\\
0 & 0 & -1 & 0\\
0 & 0 & 0 & 1
\end{array}
\right]  \ ,\tag{C6}\\
\nonumber\\
T_{3}=\frac{1}{48}\left[
\begin{array}
[c]{cccc}%
1 & 0 & 0 & 0\\
0 & -3 & 0 & 0\\
0 & 0 & 3 & 0\\
0 & 0 & 0 & -1
\end{array}
\right]  \ ,\ \ \ V^{-1}=\frac{1}{48}\left[
\begin{array}
[c]{cccc}%
-3 & 27 & 27 & -3\\
-1 & 27 & -27 & 1\\
3 & -3 & -3 & 3\\
1 & -3 & 3 & -1
\end{array}
\right]  \ . \tag{C7}%
\end{gather}

Spin $j=2$ helps to establish the general pattern.%
\begin{equation}
S^{m}=\left[
\begin{array}
[c]{ccccc}%
4^{m} & 0 & 0 & 0 & 0\\
0 & 2^{m} & 0 & 0 & 0\\
0 & 0 & 0 & 0 & 0\\
0 & 0 & 0 & \left(  -2\right)  ^{m} & 0\\
0 & 0 & 0 & 0 & \left(  -4\right)  ^{m}%
\end{array}
\right]  \ ,\ \ \ V^{-1}=\frac{1}{384}\left[
\begin{array}
[c]{ccccc}%
0 & 0 & 384 & 0 & 0\\
-16 & 128 & 0 & -128 & 16\\
-4 & 64 & -120 & 64 & -4\\
4 & -8 & 0 & 8 & -4\\
1 & -4 & 6 & -4 & 1
\end{array}
\right]  \ . \tag{C8}%
\end{equation}
From the $1$st through $5$th rows of $V^{-1}$\ one extracts, respectively, the
diagonal dual matrices $T_{0}$ through $T_{4}$.\newpage

\subsection*{Appendix D: \ A derivation using differential equations}

In this Appendix I derive the coefficients in the Cayley transform spin matrix
polynomial by solving differential equations. \ This parallels the first
derivation of the CFZ results, as given in \cite{CvK}. \ Although the
resolvent analysis for general matrices, as given in the main text, is much
simpler, perhaps the discussion here has some pedagogical value in that it
emphasizes how much more effort is needed to pursue the same differential
equation approach as was used for the exponential case.

For a given spin $j$ let $M=2i\boldsymbol{\hat{n}\cdot J}$ and write%
\begin{equation}
\frac{1}{1-\alpha M}=\sum_{m=0}^{2j}b_{m}\left(  \alpha\right)  ~M^{m}\ .
\tag{D1}\label{OriginalRationalForm}%
\end{equation}
Differentiate with respect to $\alpha$ to obtain%
\begin{equation}
\left(  1-\alpha M\right)  \frac{d}{d\alpha}\left(  \frac{1}{1-\alpha
M}\right)  =\frac{M}{1-\alpha M}\ . \tag{D2}%
\end{equation}
That is to say,%
\begin{equation}
\left(  1-\alpha M\right)  \sum_{m=0}^{2j}M^{m}\frac{db_{m}}{d\alpha}%
=\sum_{m=0}^{2j}M^{m+1}b_{m}\ , \tag{D3}%
\end{equation}
or upon rearranging the terms,%
\begin{equation}
\sum_{m=0}^{2j}M^{m}\frac{db_{m}}{d\alpha}-\sum_{m=0}^{2j-1}M^{m+1}\frac
{d}{d\alpha}\left(  \alpha b_{m}\right)  =M^{2j+1}\frac{d}{d\alpha}\left(
\alpha b_{2j}\right)  \ . \tag{D4}\label{1stOrderSeries}%
\end{equation}
Now apply the fundamental identity (\ref{FI}) to reexpress the RHS of
(\ref{1stOrderSeries}) to obtain%
\begin{equation}
\frac{db_{0}}{d\alpha}+\sum_{m=1}^{2j}M^{m}\frac{db_{m}}{d\alpha}-\sum
_{m=1}^{2j}M^{m}\frac{d}{d\alpha}\left(  \alpha b_{m-1}\right)  =-\sum
_{m=0}^{2j}M^{m}\left(  2i\right)  ^{2j+1-m}t\left(  2+2j,1+m\right)
~\frac{d}{d\alpha}\left(  \alpha b_{2j}\right)  \ . \tag{D5}%
\end{equation}
Trace projections using the dual matrices (i.e. the linear independence of the
$M^{m}$ for $0\leq m\leq2j$) gives the set of 1st order equations
\begin{align}
\frac{db_{0}}{d\alpha}  &  =-\left(  2i\right)  ^{2j+1}t\left(  2+2j,1\right)
\frac{d}{d\alpha}\left(  \alpha b_{2j}\right)  \ ,\tag{D6}\label{0thEqn}\\
\frac{db_{m}}{d\alpha}-\frac{d}{d\alpha}\left(  \alpha b_{m-1}\right)   &
=-\left(  2i\right)  ^{2j+1-m}t\left(  2+2j,1+m\right)  \frac{d}{d\alpha
}\left(  \alpha b_{2j}\right)  \text{ \ \ for }m\geq1\ . \tag{D7}%
\label{mthEqn}%
\end{align}
These are immediately integrated, with the constants of integration fixed by
the requirement that all the $b_{n}$ behave properly as $\alpha\rightarrow0$,
namely, $b_{0}\left(  \alpha=0\right)  =1$ and $b_{m}\left(  \alpha=0\right)
=0$\ for $m\geq1$. \ The only additional ingredients needed to establish
\begin{equation}
b_{n}=\frac{\alpha^{n}}{\det\left(  1-2i\alpha~\boldsymbol{\hat{n}\cdot
J}\right)  }~\operatorname*{Trunc}_{2j-n}\left[  \det\left(  1-2i\alpha
~\boldsymbol{\hat{n}\cdot J}\right)  \right]  \tag{D8}%
\label{FinalDeterminantalForm}%
\end{equation}
are $t\left(  even,odd\right)  =0=t\left(  odd,even\right)  $
\cite{Riordan,CFN} and the relations (\ref{TruncsAsCFNs}) and
(\ref{DeterminantAsCFNs}). \ The details are spelled out more fully in the
following few paragraphs.

For \emph{integer spin} (\ref{0thEqn}) and (\ref{mthEqn}) become
\begin{align}
\frac{db_{0}}{d\alpha}  &  =0\ ,\tag{D9}\label{0Integer}\\
\frac{db_{2k}}{d\alpha}  &  =\frac{d}{d\alpha}\left(  \alpha b_{2k-1}\right)
\text{ \ \ for even }m=2k\ ,\tag{D10}\label{EvenInteger}\\
\frac{db_{2k+1}}{d\alpha}  &  =\frac{d}{d\alpha}\left(  \alpha b_{2k}\right)
-4^{j-k}\left\vert t\left(  2+2j,2+2k\right)  \right\vert \frac{d}{d\alpha
}\left(  \alpha b_{2j}\right)  \text{ \ \ for odd }m=2k+1\ , \tag{D11}%
\label{OddInteger}%
\end{align}
while for \emph{semi-integer spin} the equations become%
\begin{align}
\frac{db_{0}}{d\alpha}  &  =-4^{j+\frac{1}{2}}\left\vert t\left(
2+2j,1\right)  \right\vert \frac{d}{d\alpha}\left(  \alpha b_{2j}\right)
\ ,\tag{D12}\label{0SemiInteger}\\
\frac{db_{2k}}{d\alpha}  &  =\frac{d}{d\alpha}\left(  \alpha b_{2k-1}\right)
-4^{j+\frac{1}{2}-k}\left\vert t\left(  2+2j,1+2k\right)  \right\vert \frac
{d}{d\alpha}\left(  \alpha b_{2j}\right)  \text{ \ \ for even }%
m=2k\ ,\tag{D13}\label{EvenSemiInteger}\\
\frac{db_{2k+1}}{d\alpha}  &  =\frac{d}{d\alpha}\left(  \alpha b_{2k}\right)
\text{ \ \ for odd }m=2k+1\ . \tag{D14}\label{OddSemiInteger}%
\end{align}
Here I have incorporated the phases of the central factorial numbers (see
Appendix A) to write%
\begin{align}
\left(  -1\right)  ^{j-k}t\left(  2+2j,2+2k\right)   &  =\left\vert t\left(
2+2j,2+2k\right)  \right\vert \text{ \ \ for integer }j\ ,\tag{D15}\\
\left(  i\right)  ^{2j+1-2k}t\left(  2+2j,1+2k\right)   &  =\left\vert
t\left(  2+2j,1+2k\right)  \right\vert \text{ \ \ for semi-integer }j\ .
\tag{D16}%
\end{align}
Upon integration of (\ref{EvenInteger}) and (\ref{OddSemiInteger}), with the
$\alpha=0$ initial conditions, it follows that%
\begin{align}
b_{2k}\left(  \alpha\right)   &  =\alpha b_{2k-1}\left(  \alpha\right)  \text{
\ \ for }1\leq k\leq j\text{ and integer }j\text{ ,} \tag{D17}%
\label{IntegerEvenOdd}\\
b_{2k+1}\left(  \alpha\right)   &  =\alpha b_{2k}\left(  \alpha\right)  \text{
\ \ for }0\leq k\leq j-\frac{1}{2}\text{ and semi-integer }j\ . \tag{D18}%
\label{SemiIntegerOddEven}%
\end{align}
Moreover, for integer $j$
\begin{equation}
b_{0}\left(  \alpha\right)  =1 \tag{D19}\label{0IntegerSolution}%
\end{equation}
solves (\ref{0Integer}) with the correct value at $\alpha=0$.

Equations (\ref{OddInteger}) and (\ref{EvenSemiInteger}) also integrate easily
to give two difference equations.%
\begin{align}
b_{2k+1}-\alpha b_{2k}  &  =-4^{j-k}\left\vert t\left(  2+2j,2+2k\right)
\right\vert ~\alpha b_{2j}\text{ \ \ for integer }j\ , \tag{D20}%
\label{IntegerDifference}\\
b_{2k}-\alpha b_{2k-1}  &  =-4^{j+\frac{1}{2}-k}\left\vert t\left(
2+2j,1+2k\right)  \right\vert ~\alpha b_{2j}\text{ \ \ for semi-integer
}j\text{\ .} \tag{D21}\label{SemiIntegerDifference}%
\end{align}
Note the second of these is obtained from the first by $k\rightarrow
k-\frac{1}{2}$. \ The first of these, along with (\ref{IntegerEvenOdd}), may
be solved by upward recursion starting from $k=0$, using
(\ref{0IntegerSolution}), with consistency fixing $b_{2j}=b_{2j-1}$. \ The
result for integer $j$ is%
\begin{equation}
b_{m}=\frac{\alpha^{m}}{\sum_{m=0}^{j}4^{m}\alpha^{2m}~\left\vert t\left(
2j+2,2j+2-2m\right)  \right\vert }~\sum_{k=0}^{\left\lfloor j-m/2\right\rfloor
}4^{k}\alpha^{2k}~\left\vert t\left(  2j+2,2j+2-2k\right)  \right\vert \ .
\tag{D22}\label{IntegerResultRepeated}%
\end{equation}
Note that $b_{0}=1$ when $j$ is an integer, and that $\left.  d^{m}%
b_{m}/d\alpha^{m}\right\vert _{\alpha=0}=m!$ for all $m$, a property that is
clear from the original rational form (\ref{OriginalRationalForm}).

Similarly, (\ref{SemiIntegerDifference}) may be solved by recursion to obtain
for semi-integer $j$,%
\begin{equation}
b_{m}=\frac{\alpha^{m}}{\sum_{m=0}^{\left\lfloor j+\frac{1}{2}\right\rfloor
}4^{m}\alpha^{2m}~\left\vert t\left(  2j+2,2j+2-2m\right)  \right\vert }%
~\sum_{k=0}^{\left\lfloor j-m/2\right\rfloor }4^{k}\alpha^{2k}~\left\vert
t\left(  2j+2,2j+2-2k\right)  \right\vert \ .\tag{D23}%
\label{SemiIntegerResultRepeated}%
\end{equation}
Again note that $\left.  d^{m}b_{m}/d\alpha^{m}\right\vert _{\alpha=0}=m!$ for
all $m$. \ But in this case $b_{0}\neq1$ since the $O\left(  \alpha
^{2j+1}\right)  $ term in the denominator is never present in the numerator.
\ Nevertheless, as previously announced \cite{TSvK}, in view of
(\ref{TruncsAsCFNs}) and (\ref{DeterminantAsCFNs}) it follows that the
\emph{same final} \emph{determinantal form} holds for both integer and
semi-integer $j$, namely (\ref{FinalDeterminantalForm}). \ Technically
\cite{AndAwayWeGo}, how sweet it is!

Alternatively, one may simply substitute (\ref{IntegerResultRepeated}) into
(\ref{IntegerDifference}),\ and (\ref{SemiIntegerResultRepeated})\ into
(\ref{SemiIntegerDifference}),\ to verify that those difference equations are
both satisfied, along with (\ref{IntegerEvenOdd}) and
(\ref{SemiIntegerOddEven}), with the proper behavior at $\alpha=0$.

Since for a given $j$ the $b_{n}$ are just renamed $\mathfrak{B}_{n}^{\left[
j\right]  }$, the result (\ref{GSPCoefficients})\ is proven.\newpage

\subsection*{Appendix E: \ Examples of Cayley transforms for $j=\frac{1}%
{2},\ 1,\ \frac{3}{2},\ 2,\ \frac{5}{2},$ and $3$}%

\[
\]%
\begin{equation}
\left.  \det\left(  1-2i\alpha~\boldsymbol{\hat{n}\cdot J}\right)  \right\vert
_{j=1/2}=1+\alpha^{2} \tag{E1}%
\end{equation}%
\begin{align}
\left.  \frac{1+2i\alpha~\boldsymbol{\hat{n}\cdot J}}{1-2i\alpha
~\boldsymbol{\hat{n}\cdot J}}\right\vert _{j=1/2}  &  =\frac{1}{1+\alpha^{2}%
}\left(  \left(  1-\alpha^{2}\right)  \boldsymbol{I}_{2\times2}+4i\alpha
\left(  \boldsymbol{\hat{n}\cdot J}\right)  _{2\times2}\right) \tag{E2}\\
& \nonumber
\end{align}%
\begin{equation}
\left.  \det\left(  1-2i\alpha~\boldsymbol{\hat{n}\cdot J}\right)  \right\vert
_{j=1}=1+4\alpha^{2} \tag{E3}%
\end{equation}%
\begin{align}
\left.  \frac{1+2i\alpha~\boldsymbol{\hat{n}\cdot J}}{1-2i\alpha
~\boldsymbol{\hat{n}\cdot J}}\right\vert _{j=1}  &  =\boldsymbol{I}_{3\times
3}+\frac{4i\alpha}{1+4\alpha^{2}}\left(  \boldsymbol{\left(  \boldsymbol{\hat
{n}\cdot J}\right)  }_{3\times3}+2i\alpha\left(  \boldsymbol{\hat{n}\cdot
J}\right)  _{3\times3}^{2}\right) \tag{E4}\\
& \nonumber
\end{align}%
\begin{equation}
\left.  \det\left(  1-2i\alpha~\boldsymbol{\hat{n}\cdot J}\right)  \right\vert
_{j=3/2}=\left(  1+\alpha^{2}\right)  \left(  1+9\alpha^{2}\right)
=1+10\alpha^{2}+9\alpha^{4} \tag{E5}%
\end{equation}%
\begin{align}
\left.  \frac{1+2i\alpha~\boldsymbol{\hat{n}\cdot J}}{1-2i\alpha
~\boldsymbol{\hat{n}\cdot J}}\right\vert _{j=3/2}  &  =\frac{1}{\left(
1+\alpha^{2}\right)  \left(  1+9\alpha^{2}\right)  }\left(  \left(
1+10\alpha^{2}-9\alpha^{4}\right)  \boldsymbol{I}_{4\times4}+4i\alpha\left(
1+10\alpha^{2}\right)  \left(  \boldsymbol{\hat{n}\cdot J}\right)  _{4\times
4}\right) \nonumber\\
&  +\frac{-8\alpha^{2}}{\left(  1+\alpha^{2}\right)  \left(  1+9\alpha
^{2}\right)  }\left(  \left(  \boldsymbol{\hat{n}\cdot J}\right)  _{4\times
4}^{2}+2i\alpha\left(  \boldsymbol{\hat{n}\cdot J}\right)  _{4\times4}%
^{3}\right) \tag{E6}\\
& \nonumber
\end{align}%
\begin{equation}
\left.  \det\left(  1-2i\alpha~\boldsymbol{\hat{n}\cdot J}\right)  \right\vert
_{j=2}=\left(  1+4\alpha^{2}\right)  \left(  1+16\alpha^{2}\right)
=1+20\alpha^{2}+64\alpha^{4} \tag{E7}%
\end{equation}%
\begin{align}
\left.  \frac{1+2i\alpha~\boldsymbol{\hat{n}\cdot J}}{1-2i\alpha
~\boldsymbol{\hat{n}\cdot J}}\right\vert _{j=2}  &  =\boldsymbol{I}_{5\times
5}+\frac{4i\alpha\left(  1+20\alpha^{2}\right)  }{\left(  1+4\alpha
^{2}\right)  \left(  1+16\alpha^{2}\right)  }\left(  \left(  \boldsymbol{\hat
{n}\cdot J}\right)  _{5\times5}\boldsymbol{+}2i\alpha\left(  \boldsymbol{\hat
{n}\cdot J}\right)  _{5\times5}^{2}\right) \nonumber\\
&  +\frac{-16i\alpha^{3}}{\left(  1+4\alpha^{2}\right)  \left(  1+16\alpha
^{2}\right)  }\left(  \left(  \boldsymbol{\hat{n}\cdot J}\right)  _{5\times
5}^{3}+2i\alpha\left(  \boldsymbol{\hat{n}\cdot J}\right)  _{5\times5}%
^{4}\right) \tag{E8}\\
& \nonumber
\end{align}%
\begin{equation}
\left.  \det\left(  1-2i\alpha~\boldsymbol{\hat{n}\cdot J}\right)  \right\vert
_{j=5/2}=\left(  1+\alpha^{2}\right)  \left(  1+9\alpha^{2}\right)  \left(
1+25\alpha^{2}\right)  =1+35\alpha^{2}+259\alpha^{4}+225\alpha^{6} \tag{E9}%
\end{equation}%
\begin{align}
\left.  \frac{1+2i\alpha~\boldsymbol{\hat{n}\cdot J}}{1-2i\alpha
~\boldsymbol{\hat{n}\cdot J}}\right\vert _{j=5/2}  &  =\frac{1}{\left(
1+\alpha^{2}\right)  \left(  1+9\alpha^{2}\right)  \left(  1+25\alpha
^{2}\right)  }\left(
\begin{array}
[c]{c}%
\left(  1+35\alpha^{2}+259\alpha^{4}-225\alpha^{6}\right)  \boldsymbol{I}%
_{6\times6}\\
\\
+4i\alpha\left(  1+35\alpha^{2}+259\alpha^{4}\right)  \left(  \boldsymbol{\hat
{n}\cdot J}\right)  _{6\times6}%
\end{array}
\right) \nonumber\\
&  +\frac{-8\alpha^{2}\left(  1+35\alpha^{2}\right)  }{\left(  1+\alpha
^{2}\right)  \left(  1+9\alpha^{2}\right)  \left(  1+25\alpha^{2}\right)
}\left(  \left(  \boldsymbol{\hat{n}\cdot J}\right)  _{6\times6}^{2}%
+2i\alpha\left(  \boldsymbol{\hat{n}\cdot J}\right)  _{6\times6}^{3}\right)
\nonumber\\
&  +\frac{32\alpha^{4}}{\left(  1+\alpha^{2}\right)  \left(  1+9\alpha
^{2}\right)  \left(  1+25\alpha^{2}\right)  }\left(  \left(  \boldsymbol{\hat
{n}\cdot J}\right)  _{6\times6}^{4}+2i\alpha\left(  \boldsymbol{\hat{n}\cdot
J}\right)  _{6\times6}^{5}\right) \tag{E10}\\
& \nonumber
\end{align}%
\begin{equation}
\left.  \det\left(  1-2i\alpha~\boldsymbol{\hat{n}\cdot J}\right)  \right\vert
_{j=3}=\left(  1+4\alpha^{2}\right)  \left(  1+16\alpha^{2}\right)  \left(
1+36\alpha^{2}\right)  =1+56\alpha^{2}+784\alpha^{4}+2304\alpha^{6} \tag{E11}%
\end{equation}%
\begin{align}
\left.  \frac{1+2i\alpha~\boldsymbol{\hat{n}\cdot J}}{1-2i\alpha
~\boldsymbol{\hat{n}\cdot J}}\right\vert _{j=3}  &  =\boldsymbol{I}_{7\times
7}+\frac{4i\alpha\left(  1+56\alpha^{2}+784\alpha^{4}\right)  }{\left(
1+4\alpha^{2}\right)  \left(  1+16\alpha^{2}\right)  \left(  1+36\alpha
^{2}\right)  }\left(  \left(  \boldsymbol{\hat{n}\cdot J}\right)  _{7\times
7}\boldsymbol{+}2i\alpha\left(  \boldsymbol{\hat{n}\cdot J}\right)
_{7\times7}^{2}\right) \nonumber\\
&  +\frac{-16i\alpha^{3}\left(  1+56\alpha^{2}\right)  }{\left(  1+4\alpha
^{2}\right)  \left(  1+16\alpha^{2}\right)  \left(  1+36\alpha^{2}\right)
}\left(  \left(  \boldsymbol{\hat{n}\cdot J}\right)  _{7\times7}^{3}%
+2i\alpha\left(  \boldsymbol{\hat{n}\cdot J}\right)  _{7\times7}^{4}\right)
\nonumber\\
&  +\frac{64i\alpha^{5}}{\left(  1+4\alpha^{2}\right)  \left(  1+16\alpha
^{2}\right)  \left(  1+36\alpha^{2}\right)  }\left(  \left(  \boldsymbol{\hat
{n}\cdot J}\right)  _{7\times7}^{5}+2i\alpha\left(  \boldsymbol{\hat{n}\cdot
J}\right)  _{7\times7}^{6}\right)  \tag{E12}%
\end{align}


\begin{thebibliography}{99}
\bibitem {Footnote2}At least they were unfamiliar to me when I first started
work on spin matrix expansions. \ Also, they are unnamed polynomials, as far
as I can tell.
\end{thebibliography}
\end{document}